\documentclass[12pt,twoside,english,a4paper,prd,nofootinbib,preprint,showpacs,preprintnumbers,floatfix]{revtex4}
\usepackage{mathptmx}
\usepackage{color}
\usepackage{helvet}
\usepackage{courier}
\usepackage[T1]{fontenc}
\usepackage[a4paper]{geometry}
\usepackage{units}
\usepackage{bm}
\usepackage{amsmath}
\usepackage{graphicx}
\usepackage{amssymb}

\makeatletter

\newcommand{\AcalFB}{\mathcal{A}_\text{FB}}

\makeatother

\usepackage{babel}

\begin{document}
\global\long\def\ra{\rightarrow}

\global\long\def\alphas{\alpha_{s}}

\global\long\def\ve{\varepsilon}

\global\long\def\gae{\gamma_{E}}

\global\long\def\AcalFB{{\cal A}_{FB}}

\newcommand\gsim{\mathrel{\rlap{\raise.4ex\hbox{$>$}} {\lower.6ex\hbox{$\sim$}}}}  \newcommand\lsim{\mathrel{\rlap{\raise.4ex\hbox{$<$}} {\lower.6ex\hbox{$\sim$}}}}  

\newcommand\etmiss{E_{T}\hspace{-13pt}/\hspace{8pt}}

\newcommand{\mHQ}{m_{\mathcal Q}} 

\newcommand{\SLASH}[2]{\makebox[#2ex][l]{$#1$}/} \newcommand{\Dslash}{\SLASH{D}{.5}\,} \newcommand{\kslash}{\SLASH{k}{.15}} 
\newcommand{\nslash}{\SLASH{n}{.15}} 
\newcommand{\pslash}{\SLASH{p}{.2}} 
\newcommand{\qslash}{\SLASH{q}{.08}} \newcommand{\syslash}{\SLASH{s}{.06}} \newcommand{\aslash}{\SLASH{a}{.06}} \newcommand{\bslash}{\SLASH{b}{.06}} \newcommand{\parsl}{\SLASH{\partial}{.1}}

\newcommand{\Mcal}{\mathcal{M}}

%
%

\begin{flushright}
MZ-TH/12-09\\
TTP12-005\\
\par\end{flushright}

\vspace{4mm}

\begin{center}
\textbf{\LARGE MSSM corrections to the top-antitop quark}
\par\end{center}{\LARGE \par}

\begin{center}
\textbf{\LARGE forward-backward asymmetry at the Tevatron }\vspace{7mm}

\par\end{center}

\begin{center}
\textbf{\large Stefan Berge}$^{*}$\textbf{\large }%
\footnote{\texttt{\small berge@uni-mainz.de}%
}\textbf{\large{} , Doreen Wackeroth}$^{\$}$\textbf{\large }%
\footnote{\texttt{\small dow@ubpheno.physics.buffalo.edu}%
}\textbf{\large , Martin Wiebusch}$^{\dagger}$\textbf{\large }%
\footnote{\texttt{\small wiebusch@particle.uni-karlsruhe.de}%
}
\par\end{center}{\large \par}

\begin{center}
$^{*}$ Institute for Physics (WA THEP), Johannes Gutenberg-Universit\"at,
\\
D-55099 Mainz, Germany\vspace{-7mm}

\par\end{center}

\noindent \begin{center}
$^{\$}$ Department of Physics, University at Buffalo, The State University
of New York, \\
Buffalo, NY 14260-1500, U.S.A.\vspace{-3mm}

\par\end{center}

\begin{center}
$^{\dagger}$ Institute for Theoretical Particle Physics, Karlsruhe
Institute of Technology (KIT), D-76128 Karlsruhe, Germany
\par\end{center}

\begin{center}
\vspace{12mm}
 \textbf{Abstract}
\par\end{center}

We study the effects of the complete supersymmetric QCD and
electroweak one-loop corrections to the $t\bar{t}$ forward-backward
asymmetry at the Fermilab Tevatron $p\bar{p}$ collider. We work in the
complex Minimal Supersymmetric Standard Model (MSSM), only restricted
by the condition of minimal flavor violation (MFV).  We perform a
comprehensive scan over the relevant parameter space of the complex
MFV-MSSM and determine the maximal possible contributions of these
MSSM loop corrections to the forward-backward asymmetry in the $t\bar
t$ center-of-mass frame.

\vfill{}

\newpage{}

%
\section{Introduction}\label{sec:intro}
%

Precision studies of top quark properties at the Fermilab Tevatron
$p\bar p$ and CERN LHC $pp$ colliders keep probing the Standard Model
(SM) of electroweak (EW) and strong interactions at an increasing
level of precision, and may provide a window to new physics. While the
total $t\bar t$ cross section and $t\bar t$ invariant mass ($M_{t \bar
t}$) distribution agree with SM predictions within their respective
uncertainties, a measurement of the corrected (parton-level)
forward-backward asymmetry in top-pair production,
$A_{\text{FB}}^{t\bar t}$, by the
CDF~\cite{Aaltonen:2008hc,Aaltonen:2011kc,unknown:2011comb} and
D0~\cite{:2007qb,Abazov:2011rq} collaborations at the Tevatron
\begin{eqnarray}
A_{\text{FB}}^{t\bar t}(\mbox{CDF~\cite{unknown:2011comb}})&=&
  20 \pm 7_\text{stat} \pm 2_\text{syst}\% \nonumber \\
A_{\text{FB}}^{t\bar t}(\mbox{D0~\cite{Abazov:2011rq}})&=& 19.6 \pm 6.5 \% 
\end{eqnarray}
differs by about $2\sigma$~\cite{unknown:2011comb,Abazov:2011rq} from
the SM QCD prediction.  The difference between measurement and SM QCD
prediction is even more pronounced, i.~e. at the $3\sigma$ level, in
the region $M_{t\bar t}>450$~GeV, where a measurement of
$A_{\text{FB}}^{t\bar t}$ yields~\cite{Aaltonen:2011kc}:
\begin{equation}
A_{\text{FB}}^{t\bar t}(\mbox{CDF},M_{t\bar t}>450~\mbox{GeV})= 47.5 \pm 11.4 \%
\end{equation}
The forward-backward asymmetry in the $t\bar t$ center-of-mass (CM) frame
is defined as:
\begin{equation}
A_{\text{FB}}^{t\bar t}=\frac{\sigma_{t\bar t}(\Delta y >0)-\sigma_{t\bar t}(\Delta y <0)}
{\sigma_{t\bar t}(\Delta y >0)+\sigma_{t\bar t}(\Delta y <0)}
\end{equation}
where $\Delta y=y_t-y_{\bar t}$ denotes the difference in rapidity
of the top and anti-top quark, and SM predictions including higher-order QCD and
EW corrections are provided in
Refs.~\cite{Kuhn:1998kw,Kuhn:1998jr,Bowen:2005ap,Antunano:2007da,Almeida:2008ug,Ahrens:2011uf,Hollik:2011ps,Kuhn:2011ri}
and Refs.~\cite{Hollik:2011ps,Kuhn:2011ri,Manohar:2012rs}, respectively.  The
interpretation of the observed discrepancy requires a solid understanding of the
theory predictions, i.~e. control of the theoretical uncertainties. Recent
updated calculations and studies of theoretical uncertainties at next-to-leading
order (NLO) and next-to-next-to-leading-logarithmic order (NNLL) in
QCD~\cite{Kidonakis:2011zn,Ahrens:2011uf} and at NLO
EW+QCD~\cite{Hollik:2011ps,Kuhn:2011ri} find that the discrepancy is reduced
compared to NLO QCD predictions, but still persists for the measurement at large
$M_{t\bar t}$. For instance, including the EW contributions to ${\cal
  O}(\alpha^2)$ and ${\cal O}(\alpha \alpha_s^2)$ to $A_{\text{FB}}^{t\bar t}$
results in a combined NLO QCD+EW prediction of~\cite{Hollik:2011ps} (including
the factorization/renormalization scale uncertainty and using MRST2004QED)
\begin{eqnarray}
A_{\text{FB}}^{t\bar t}(\mbox{NLO QCD+EW})&=& 8.93^{+0.79}_{-0.62} \% \nonumber \\ 
A_{\text{FB}}^{t\bar t}(\mbox{NLO QCD+EW},M_{t\bar t}>450~\mbox{GeV})&=& 12.77^{+1.13}_{-0.86} \% \, , 
\end{eqnarray}
and when including NNLL QCD contributions $A_{\text{FB}}^{t\bar t}$ is
predicted as~\cite{Ahrens:2011uf} (including the
factorization/renormalization scale uncertainty and using MSTW2008):
\begin{eqnarray}
A_{\text{FB}}^{t\bar t}(\mbox{NLO+NNLL QCD})&=& 7.24^{+1.04}_{-0.67} \%  \nonumber \\ 
A_{\text{FB}}^{t\bar t}(\mbox{NLO+NNLL QCD},M_{t\bar t}>450~\mbox{GeV})&=& 11.1^{+1.7}_{-0.9} \% \, . 
\end{eqnarray}
However, since the first non-vanishing contribution to
$A_{\text{FB}}^{t\bar t}$ is of NLO in QCD in the $t\bar t$ production
cross section, a conclusive answer concerning the theoretical
uncertainty will only be possible once a calculation of the complete
NNLO QCD corrections to $t\bar t$ production becomes
available. Nevertheless, it is interesting to study the possibility
that the observed discrepancy could be interpreted as a signal of new
physics. Possible SM extensions which may give rise to large
contributions to $A_{\text{FB}}^{t\bar t}$ have been explored
extensively in the literature and some recent examples can be found in
Refs.~\cite{AguilarSaavedra:2011ug,Davoudiasl:2011tv,Cui:2011xy,Isidori:2011dp}.
In this paper, we consider the one-loop ${\cal O}(\alpha_s)$ SUSY QCD
and ${\cal O}(\alpha)$ SUSY EW corrections to the strong partonic
$t\bar t$ production process, $q\bar q \to t\bar t$, and study their
impact on $A_{\text{FB}}^{t \bar t}$ at the Tevatron.  We work in the
complex Minimal Supersymmetric SM (MSSM)
~\cite{Nilles:1983ge,Haber:1984rc} and assume it to only be
restricted by the condition of minimal flavor violation 
(MFV)~\cite{Buras:2000dm,D'Ambrosio:2002ex}.
The other possible partonic process at leading-order (LO) QCD, $gg\to
t\bar t$, is symmetric in the production rates for top quarks in the
forward and backward hemisphere, and thus only enters the total cross
section in the denominator of $A_{\text{FB}}^{t \bar t}$. After a
consistent perturbative expansion of $A_{\text{FB}}^{t \bar t}$ in
$\alpha_s$ and $\alpha$, the MSSM one-loop contribution to
$A_{\text{FB}}^{t \bar t}$ calculated in this paper can then be
written as follows (the dependence on $\alpha_s$ and $\alpha$ is
explicitly shown):\vspace*{6pt}
\begin{equation}\label{eq:main}
A_{\text{FB}}^{t\bar t}=\alpha_s \frac{\Delta \sigma_{t\bar t}^{SQCD}}{\sigma_{t\bar t}^{(0)}} 
+\alpha \frac{\Delta\sigma_{t\bar t}^{SEW}}{\sigma_{t\bar t}^{(0)}} 
\end{equation}\\[2pt]
with $\Delta \sigma_{t\bar t}^{SQCD,SEW}=\delta\sigma_{t\bar
t}^{SQCD,SEW}(\Delta y >0)-\delta\sigma_{t\bar t}^{SQCD,SEW}(\Delta y
<0)$, where $\sigma_{t\bar t}^{(0)}$ denotes the total $t\bar t$
production cross section at LO QCD and $\delta\sigma_{t\bar
t}^{SQCD,SEW}$ denote the SUSY QCD and SUSY EW one-loop contributions,
respectively.  The SUSY QCD and SUSY EW one-loop corrections to
(unpolarized) $t\bar t$ production in hadronic collisions have been
studied in
Refs.~\cite{Li:1995fj,Alam:1996mh,Sullivan:1996ry,Zhou:1997fw,Yu:1998xv,Wackeroth:1998wm,Berge:2007dz,Ross:2007ez}
and
Refs.~\cite{Yang:1995hq,Yang:1996dm,Kim:1996nz,Hollik:1997hm,Ross:2007ez},
respectively. They are known to only modestly impact the total $t\bar
t$ production cross section and invariant $t\bar t$ mass distribution,
and thus, at the presently available precision, do not spoil the good
agreement between theory and experiment for those observables, at
least for sparticle masses of ${\cal O}(100)$~GeV (and larger) and
$m_{\tilde g}  \gsim 230$~GeV~\cite{Berge:2007dz}.  In this paper, we will derive general bounds
on $A_{\text{FB}}^{t\bar t}$, i.~e. we work within the MFV-MSSM and scan over
a large range of values for the relevant MSSM input parameters,
without imposing additional constraints. In particular, these bounds
do not rely, for instance, on specific SUSY breaking scenarios or an
artificially reduced parameter space such as the constrained MSSM
(CMSSM). As will be discussed as well, the impact of any specific
assumption, e.~g., mass limits derived from LHC squark and gluino
searches, on the bounds on $A_{\text{FB}}^{t\bar t}$ can then be readily
deduced from these general results.

The paper is organized as follows. After presenting the calculation of
the MSSM one-loop corrections that contribute to the forward-backward
asymmetry in Section~\ref{sec:MSSM-Contributions-to_Afb}, we derive
analytic expressions for bounds on $A_{\text{FB}}^{t \bar t}$ induced
by SUSY QCD one-loop corrections in Section~\ref{sec:sqcd} and discuss
the structure of the SUSY EW one-loop corrections to $A_{\text{FB}}^{t
\bar t}$ in more detail in Section~\ref{sec:sew}.  In
Section~\ref{sec:results}, we present numerical results for the
forward-backward asymmetry. In Section~\ref{sec:results:loop}, we
determine general bounds on $A_{\text{FB}}^{t\bar t}$ based on the analytic
expressions derived in Section~\ref{sec:sqcd}.  In
Section~\ref{sec:results:mssm}, we present a comprehensive scan over
the relevant complex MFV-MSSM parameter space.  We conclude in
Section~\ref{sec:conclusions} and provide explicit expressions for the
relevant MSSM couplings and one-loop corrections in the appendix.

%
\section{MSSM One-loop Contributions to $\bm{A_{\text{FB}}^{t\bar t}}$
\label{sec:MSSM-Contributions-to_Afb}}
%

\begin{figure}[h]
\includegraphics[scale=0.5]{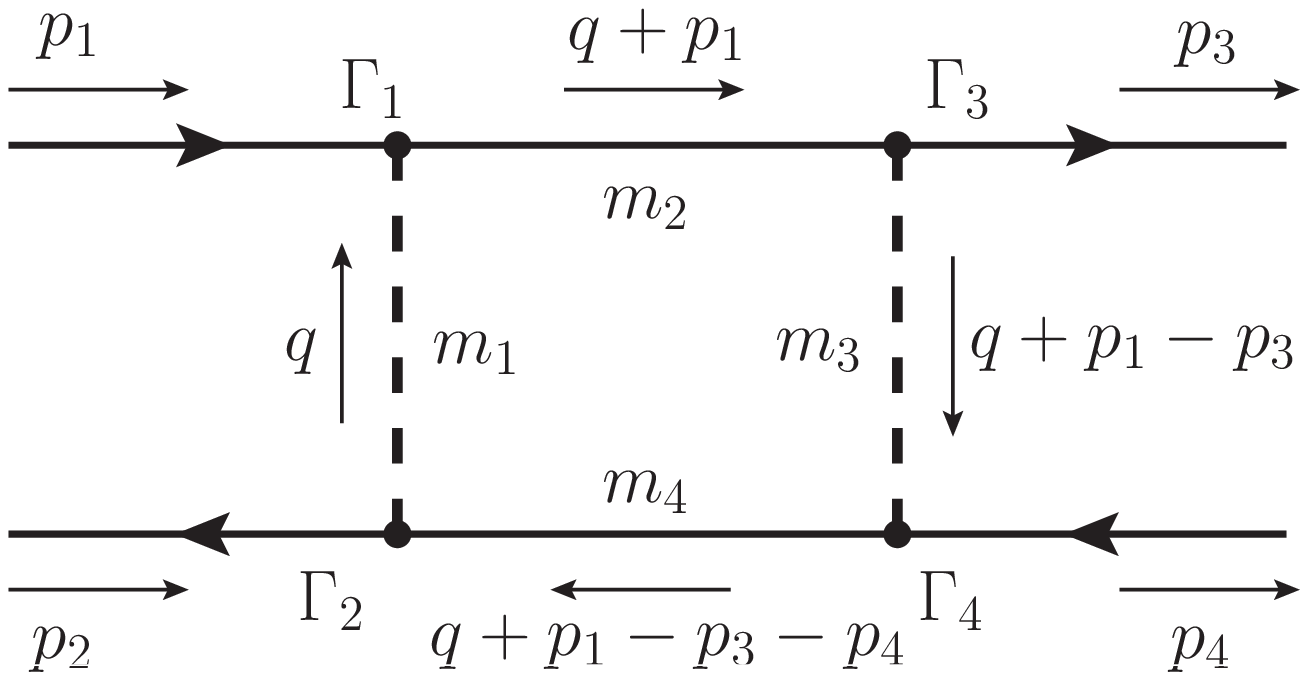}\hspace*{1cm}\includegraphics[scale=0.5]{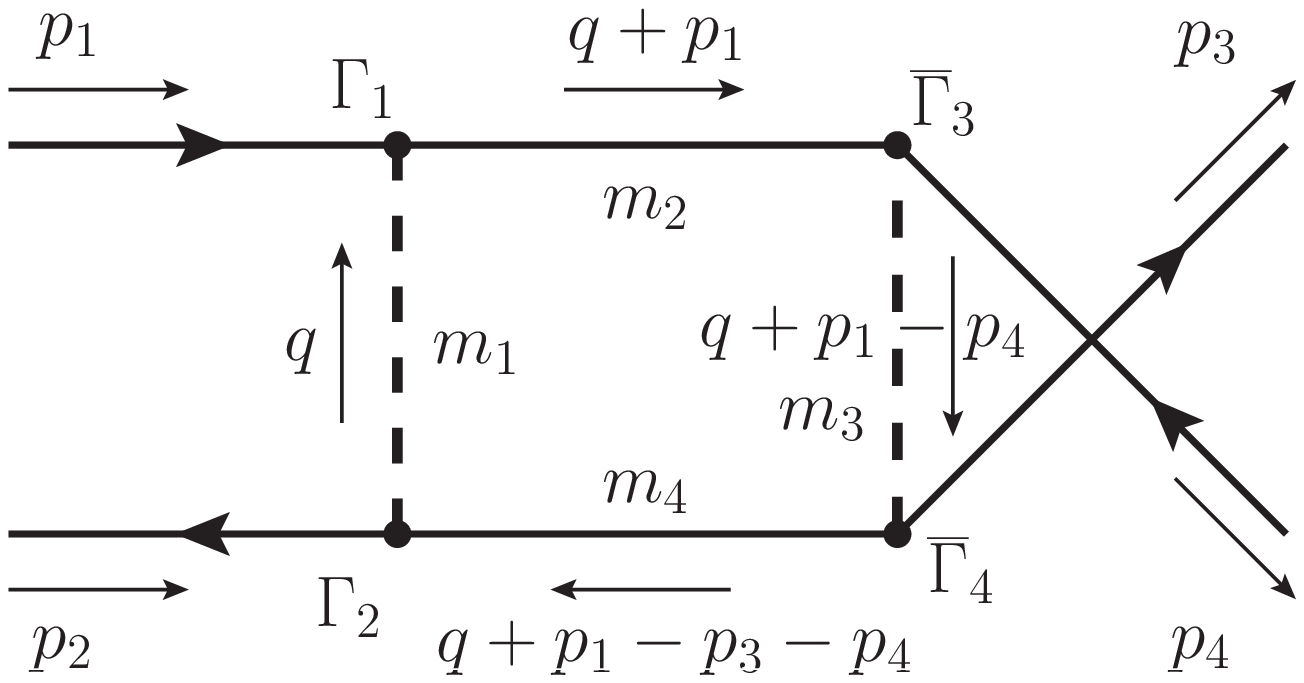}

\caption{Momentum and mass assignments for the direct and crossed box diagrams.
A possible color factor has been omitted.\label{fig:boxes}}
\end{figure}

In the MSSM the additional one-loop contributions to the top
forward-backward asymmetry, i.~e.  $A_{\text{FB}}^{t\bar t}$ of
Eq.~\eqref{eq:main}, originate from box diagrams involving squarks,
gluinos and neutralinos inside the loops. The generic diagrams are
shown in Fig.~\ref{fig:boxes}.  The external momenta are denoted by
$p_{1},\ldots,p_{4}$ and $m_{1},\ldots,m_{4}$ are the masses of the
internal particles. The gluino-squark-quark and
neutralino-squark-quark vertices are represented by
$\Gamma_{n}$ ($n=1,\ldots,4$), which we decompose, excluding a possible
color factor, in the following way:\vspace*{4pt}
\begin{eqnarray}
\Gamma_{n} & = & g_{n}^{+}P^{+}+g_{n}^{-}P^{-}\qquad{\rm with}\qquad P^{\pm}=\frac{1\pm\gamma_{5}}{2}\quad.
\label{eq:Def_Gamma_n}
\end{eqnarray}\\[-13pt]
The full gluino-squark-quark vertex also includes color matrices
$\Gamma_{n}^{c}=T^{c}\cdot\Gamma_{n}$ with $T^{c}=\lambda^{c}/2$ and
$\lambda^{c}$ the Gell-Mann matrices. The internal masses $m_{n}$ and
the generic couplings $g_{n}^{\pm}$ depend on the particles that
are assigned to the internal lines. Explicit
expressions for $g_{n}^{\pm}$ for gluino-squark-quark and
neutralino-squark-quark couplings are provided in
Section~\ref{sec:gsqcoup} and Section~\ref{sec:nsqcoup}, respectively.
For each assignment of internal particles to the direct box
(Fig.~\ref{fig:boxes}a), we get a corresponding contribution from the
crossed box (Fig.~\ref{fig:boxes}b) with identical values of the
internal masses $m_{n}$ and the coupling parameters $g_{n}^{\pm}$. The
vertices $\overline{\Gamma}_{n}$ ($n=3,4$) are related to the
$\Gamma_{n}$ by\vspace*{4pt}
\[
\overline{\Gamma}_{n}=\gamma^{0}\Gamma_{n}^{\dagger}\gamma^{0}=g_{n}^{-*}P^{+}+g_{n}^{+*}P^{-}.\]
\vspace*{-11pt}

The amplitudes ${\cal M}^{\text{(a)}}$ and ${\cal M}^{\text{(b)}}$
corresponding to the diagrams in Fig.~\ref{fig:boxes} can be generated
using the Feynman rules for fermion number violating interactions of
Ref.~\cite{Denner:1992vz} that have been implemented in
  FeynArts~\cite {Kublbeck:1990xc,Hahn:2000kx,Hahn:2001rv}.
They are then calculated with standard
trace techniques and a  
Passarino-Veltman reduction~\cite{Passarino:1979jh} of the loop
integrals
using FORM~\cite{Vermaseren:2000nd}.
For the direct box diagram we write the partonic
differential cross section $\frac{d\hat{\sigma}^{(a)}}{d\cos\theta}$
in terms of the interference of ${\cal M}^{\text{(a)}}$ with the LO
matrix element ${\cal M}^{(0)}$ as\vspace*{5pt}
\begin{eqnarray}
\frac{d\hat{\sigma}^{(a)}}{d\cos\theta}&=&\frac{\beta_t}{32\pi\hat{s}}\cdot
\smash{\frac{1}{4}\cdot\frac{1}{9}\sum_{\text{spins}}\!2Re\left\{ {\cal M}^{\text{(a)}}{\cal M}^{(0)*}\right\} } \nonumber \\[8pt]
& = & \frac{\beta_t}{32\pi\hat{s}}\cdot \smash{N_{a}\!\sum_{\lambda\,=\,\pm}\!\,2\, Re\!\left[\,\,\,\, g_{1}^{+\lambda}g_{2}^{-\lambda}g_{3}^{+\lambda}g_{4}^{+\lambda}D^{+-++}(\hat{s},\cos\theta)\right.}\phantom{{\left[\quad g_{1}^{+1}\right.}}\nonumber \\
 &  & \hspace*{3cm}{}+\, g_{1}^{-\lambda}g_{2}^{+\lambda}g_{3}^{+\lambda}g_{4}^{+\lambda}D^{-+++}(\hat{s},\cos\theta)\phantom{{\left[\quad g_{1}^{+1}\right.}}\nonumber \\
 &  & \hspace*{3cm}{}+\, g_{1}^{-\lambda}g_{2}^{+\lambda}g_{3}^{+\lambda}g_{4}^{-\lambda}D^{-++-}(\hat{s},\cos\theta)\phantom{{\left[\quad g_{1}^{+1}\right.}}\nonumber \\
 &  &
 \hspace*{3cm}{}+\smash{\left.g_{1}^{-\lambda}g_{2}^{+\lambda}g_{3}^{-\lambda}g_{4}^{+\lambda}D^{-+-+}(\hat{s},\cos\theta)
     \,\,\right]}\phantom{{\left[\quad
       g_{1}^{+1}\right.}}\label{eq:intf_a}
\end{eqnarray}\\[-10pt]
with $\beta_t=\sqrt{1-\frac{4m_{t}^{2}}{\hat{s}}}$ and
$m_t$ denoting the top quark mass. The sum over the index $\lambda$
has to be
interpreted as\vspace*{3pt}
\[\sum_{\lambda=\pm} g^{\pm \lambda}_n = \sum_{\lambda={+1,-1}}
g^{(\pm 1)*(\lambda)}_n= g^{\pm}_n + g^{\mp}_n\]\\[3pt]
with $g^{\pm}_n$ defined in Eq.~\eqref{eq:def_g_gluino}
for SUSY QCD and in  Eq.~\eqref{eq:def_g_Neut}
for SUSY EW loop diagrams.
The factors $1/4$ and $1/9$  in Eq.~\eqref{eq:intf_a}
are the spin and color average factors, respectively. The
functions $D^{a}(\hat{s},\cos\theta)$ with
 \begin{eqnarray}
a & \in & \{+-++,-+++,-++-,-+-+\}
\label{eq:pm_combinations}
\end{eqnarray}\\[-12pt]
are given in Eq.~\eqref{eq:Definition_of_Da}. They depend on the
partonic Mandelstam variables
$\hat{s}=(p_{1}+p_{2})^{2}=(p_{3}+p_{4})^{2}$ and by
$\hat{t}=(p_{1}-p_{3})^{2}=(p_{2}-p_{4})^{2}$ on the cosine of the
scattering angle $\theta$, i.e.\ the angle between the spatial
components of $p_{1}$ and $p_{3}$. They also depend on the internal
masses $m_{i}$, but these arguments are suppressed in
Eq.~\eqref{eq:intf_a} and the following discussion.  Also, in
Eq.~\eqref{eq:intf_a} (and in the following discussion) it is
implicitly understood that one has to sum over all possible
assignments of particles to the internal lines.  Note that the
functions $D^{a}(\hat{s},\cos\theta)$ do not depend on the generic
couplings $g_{n}^{\pm}$. The factor $N_{a}$ is the color factor and
contains the factor $1/9$ from the color average. The factor $1/4$
from the spin average is absorbed into the functions
$D^a(\hat{s},\cos\theta)$.

The interference of the crossed box (Fig.~\ref{fig:boxes}b) with
${\cal M}^{(0)}$ can also be expressed in terms of the functions
$D^{a}(\hat{s},\cos\theta)$. To see this, we note that the crossed box
is obtained from the direct box by interchanging the momenta $p_{3}$
and $p_{4}$, reversing the fermion flow on the outgoing legs and
replacing $\Gamma_{3}$ and $\Gamma_{4}$ by $\bar{\Gamma}_{3}$ and
$\bar{\Gamma}_{4}$, respectively. The interchange of $p_{3}$ and
$p_{4}$ is achieved by replacing $\cos\theta$ by $-\cos\theta$.  When
calculating the interference with the LO diagram, the appearance of
fermion-number violating vertices must be handled correctly, e.~g.\
following the rules in Ref.~\cite{Denner:1992vz}, and leads to an
overall minus sign. The replacements $\Gamma_{3}\to\bar{\Gamma}_{3}$
and $\Gamma_{4}\to\bar{\Gamma}_{4}$ are equivalent to $g_{n}^{\pm}\to
g_{n}^{\mp*}$ ($n=3,4$). Thus, we obtain\vspace*{5pt}
\begin{eqnarray}
\frac{d\hat{\sigma}^{(b)}}{d\cos\theta}&=&\frac{\beta_t}{32\pi\hat{s}}\cdot
\smash{\frac{1}{4}\cdot\frac{1}{9}\sum_{\text{spins}}\!2Re\left\{ {\cal M}^{\text{(b)}}{\cal M}^{(0)*}\right\} } \nonumber \\[8pt]
& = & \frac{\beta_t}{32\pi\hat{s}}\cdot N_{b}\!\smash{\sum_{\lambda=+,-}}\!\!\!2\, Re\!\bigl[{}-g_{1}^{+\lambda}g_{2}^{-\lambda}g_{3}^{-\lambda*}g_{4}^{-\lambda*}D^{+-++}(\hat{s},-\cos\theta)\nonumber \\
 & & \hspace*{3.3cm}{} -g_{1}^{-\lambda}g_{2}^{+\lambda}g_{3}^{-\lambda*}g_{4}^{-\lambda*}D^{-+++}(\hat{s},-\cos\theta)\nonumber \\
 & & \hspace*{3.3cm}{}-g_{1}^{-\lambda}g_{2}^{+\lambda}g_{3}^{-\lambda*}g_{4}^{+\lambda*}D^{-++-}(\hat{s},-\cos\theta)\nonumber \\
 &&  \hspace*{3.3cm}{}-g_{1}^{-\lambda}g_{2}^{+\lambda}g_{3}^{+\lambda*}g_{4}^{-\lambda*}D^{-+-+}(\hat{s},-\cos\theta)\,\,\bigr].\label{eq:intf_b}
\end{eqnarray}\\[-12pt]
with $D^a$ of Eq.~\eqref{eq:Definition_of_Da}.  Note that the
interference of the crossed box with the LO matrix element ${\cal
M}^{(0)}$ may have a different color factor than the direct box. This
factor is denoted by $N_{b}$.

Using Eqs.~\eqref{eq:intf_a},~\eqref{eq:intf_b} and $d\hat \sigma=d\hat
\sigma^{(a)}+d\hat \sigma^{(b)}$, the corresponding partonic
forward-backward asymmetry can be written in a compact form ($\hat
\sigma_{t\bar t}^{(0)}$ denotes the partonic LO total $t\bar t$ cross
section, including both the $q\bar q$ and $gg$-initiated $t\bar t$
production processes)\vspace*{7pt}
\begin{eqnarray}
\hat{A}_{FB}^{t\bar t}(\hat s) & = & \frac{1}{\hat \sigma_{t \bar t}^{(0)}}
\int_0^1 d\cos(\theta) \frac{d\hat{\sigma}(\hat{s},\cos\theta)}{d\cos\theta}-\int_{-1}^0 d\cos(\theta) \frac{d\hat{\sigma}(\hat{s},\cos\theta)}{d\cos\theta} \\ \nonumber
&=&
\frac{1}{\hat \sigma_{t \bar t}^{(0)}}
\int_0^1 d(\cos\theta) [\frac{d\hat{\sigma}(\hat{s},\cos\theta)}{d\cos\theta}-\frac{d\hat{\sigma}(\hat{s},-\cos\theta)}{d\cos\theta}] \\ \nonumber
&=&
\frac{1}{\hat \sigma_{t \bar t}^{(0)}}
\int_0^1 d(\cos\theta) \sum_{a}^{\,^{\,}}\, Re\left\{ \, G^{a}_q\cdot\hat{A}^{a}(\hat{s},\cos\theta)\,\right\} 
\end{eqnarray}\\[-3pt]
with the index $a$ defined in Eq.~\eqref{eq:pm_combinations}, \vspace*{3pt}
\begin{eqnarray}
\hat{A}^{a}(\hat{s},\cos\theta) & = & \frac{\beta_t}{32\pi\hat{s}}\cdot N_g\cdot\left[2\, D^{a}(\hat{s},\cos\theta)-2\, D^{a}(\hat{s},-\cos\theta)\right]\label{eq:Def_Aa}
\end{eqnarray}
and 
\begin{eqnarray}
G^{+-++}_q & =\smash{\sum_{\lambda=+,-}}\hat g_{1}^{+\lambda}\hat g_{2}^{-\lambda}(\hat g_{3}^{+\lambda}\hat g_{4}^{+\lambda}N_{a}+\hat g_{3}^{-\lambda*}\hat g_{4}^{-\lambda*}N_{b}),\nonumber \\
G^{-+++}_q & =\smash{\sum_{\lambda=+,-}}\hat g_{1}^{-\lambda}\hat g_{2}^{+\lambda}(\hat g_{3}^{+\lambda}\hat g_{4}^{+\lambda}N_{a}+\hat g_{3}^{-\lambda*}\hat g_{4}^{-\lambda*}N_{b}),\nonumber \\
G^{-++-}_q & =\smash{\sum_{\lambda=+,-}}\hat g_{1}^{-\lambda}\hat g_{2}^{+\lambda}(\hat g_{3}^{+\lambda}\hat g_{4}^{-\lambda}N_{a}+\hat g_{3}^{-\lambda*}\hat g_{4}^{+\lambda*}N_{b}),\nonumber \\
G^{-+-+}_q & =\smash{\sum_{\lambda=+,-}}\hat g_{1}^{-\lambda}\hat g_{2}^{+\lambda}(\hat g_{3}^{-\lambda}\hat g_{4}^{+\lambda}N_{a}+\hat g_{3}^{+\lambda*}\hat g_{4}^{-\lambda*}N_{b}) \, .
\label{eq:Gi}
\end{eqnarray}
Note that the coupling products $G^{a}_q$ may, in general, depend on
the initial-state quark flavor $q$, as indicated by the subscript. The
color factors $N_{a,b}$ have been absorbed into the definition of
$G^{a}_q$ and all the other factors into the definition of
$\hat{A}^{a}(\hat{s},\cos\theta)$.  An additional coupling factor
$N_g$ has been absorbed in $\hat A^{a}$ for convenience so that the
coupling products $G_{q}^{a}$, defined in terms of $\hat g_n^{\pm}$ of
Eqs.~\eqref{eq:def_hat_g_gluino} and \eqref{eq:def_hat_g_Neut}, are of ${\cal
O}(1)$.

The hadronic forward-backward asymmetry is then defined by folding
with the parton distribution functions (PDF) and by dividing by the
total hadronic LO cross section $\sigma_{t \bar t}^{(0)}$, so that
$A_{\text{FB}}^{t \bar t}$ of Eq.~\eqref{eq:main} at the Tevatron
reads\vspace*{5pt}
\begin{eqnarray}\label{eq:afbtt}
A_{\text{FB}}^{t\bar t}&=&\frac{1}{\sigma_{t\bar t}^{(0)}}
\sum_q \int_{0}^{1}dx_{1}\int_{0}^{1}dx_{2} \int_0^1 d(\cos\theta) \nonumber \\
&\times&  \left[\,\,\,\, f_{q/p}(x_{1})f_{\bar q/\bar p}(x_{2})
 \sum_{a}^{\,^{\,}} \, Re\left\{ \, G^{a}_q\cdot\hat{A}^{a}(\hat{s},\cos\theta) \,\right\} \theta(\hat{s}-4m_{t}^{2}) \right.\nonumber \\
&& \,\,\,\, {}+\left. f_{\bar q/p}(x_{1})f_{q/\bar p}(x_{2})
 \sum_{a}^{\,^{\,}}\, Re\left\{ \, G^{a}_q\cdot\hat{A}^{a}(\hat{s},-\cos\theta) \,\right\} \theta(\hat{s}-4m_{t}^{2}) \right]_{\hat s=x_1 x_2 S}  \nonumber \\
&=& \sum_q Re\left\{\sum_a \,  G^a_q A_q^a \right\} = \sum_q A_q  \, ,
\end{eqnarray}\\[-5pt]
where we defined for each initial-state quark flavor $q$\vspace*{5pt}
\begin{eqnarray}
A_{q}^{a}&=&\frac{1}{\sigma_{t\bar t}^{(0)}}\int_{0}^{1}dx_{1}\int_{0}^{1}dx_{2}\int_{0}^{1}d(\cos\theta)[f_{q/p}(x_{1})f_{q/p}(x_{2})-f_{\bar{q}/p}(x_{1})f_{\bar{q}/p}(x_{2})] \nonumber \\
& \times& \theta(\hat{s}-4m_{t}^{2})\cdot\hat{A}^{a}(\hat{s},\cos\theta)\Bigr|_{\hat{s}=x_{1}x_{2}S}
\label{eq:hadron_AFB_a}
\end{eqnarray}
and
\begin{align}\label{eq:Hadron_AFBq}
  A_q = \smash Re\bigl\{
  &\phantom{{}+{}}G_{q}^{+-++}A_{q}^{+-++}\,\,+\,\, G_{q}^{-+++}A_{q}^{-+++}
  \nonumber\\
  &+G_{q}^{-++-}A_{q}^{-++-}\,\,+\,\, G_{q}^{-+-+}A_{q}^{-+-+}\bigr\} \, .
\end{align}\\[-5pt]
Here $\sqrt{S}=\unit{1.96}$~TeV is the Tevatron hadronic CM
  energy. The function $f_{q/p}(f_{\bar q/p})$ is the PDF of the
  quark(anti-quark) flavor $q(\bar q)$ inside the proton and the
  functions $f_{q/\bar p}$ and $f_{\bar q/\bar p}$ are the
  corresponding PDFs for the anti-proton.  In
  Eq.~\eqref{eq:hadron_AFB_a} we made use of the fact that at a $p\bar
  p$ collider the quark(anti-quark) distribution inside the proton
  coincides with the anti-quark(quark) distribution inside the
  anti-proton. Moreover, when the incoming momenta are interchanged
  (or $x_1 \leftrightarrow x_2$), it corresponds to replacing
  $\cos\theta$ with $-\cos\theta$. The difference in the PDFs occurs
  because $\hat A^{a}$ of Eq.~\eqref{eq:Def_Aa} is anti-symmetric in
  $\cos\theta$.

In the presence of $CP$ violating phases, the coupling
factors $G_{q}^{a}$ of Eq.~\eqref{eq:Gi} 
may have imaginary parts. In this case, the
imaginary parts of the $A_q^{a}$ may also contribute to
$A_{\text{FB}}^{t\bar t}$. 

With the notations above, we have decomposed the MSSM one-loop
contribution to the forward-backward asymmetry, $A_{\text{FB}}^{t\bar
t}$ of Eq.~\eqref{eq:main}, into loop functions $A_{q}^{a}$ and coupling
products $G_{q}^{a}$. The loop functions only depend on the internal
masses in the box diagrams of Fig.~\ref{fig:boxes} while the
dependence on the coupling constants is contained in the coupling
factors $G_{q}^{a}$.  This separation will prove very useful when we
attempt to give bounds on $A_{\text{FB}}^{t\bar t}$ in the
MSSM that do not rely on specific SUSY breaking scenarios or an
artificially reduced parameter space.

%
\subsection{SUSY QCD One-loop Contributions\label{sec:sqcd}}
\begin{figure}[h]
\includegraphics[scale=0.6]{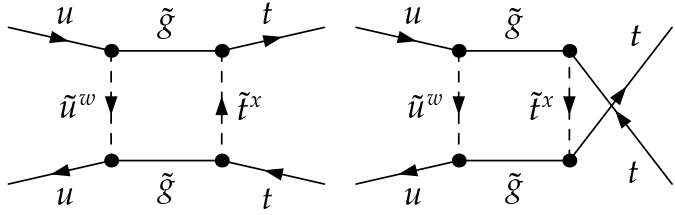}

\caption{Feynman diagrams of the SUSY QCD one-loop contribution to the forward-backward asymmetry $A_{\text{FB}}^{t\bar t}$.
\label{fig:Feynman_Diags_SQCD}}
\end{figure}

When only SUSY QCD one-loop contributions to $A_{\text{FB}}^{t\bar t}$
are considered, the Majorana fermions in the box diagrams of
Fig.~\ref{fig:boxes} are all gluinos. The relevant diagrams are shown
in Fig.~\ref{fig:Feynman_Diags_SQCD}.
In the case of SUSY QCD, Eq.~\eqref{eq:Hadron_AFBq} can be further
simplified due to relations among the coupling products $G_{q}^{a}$
and the loop functions $A_{q}^{a}$. The color factors in
Eq.~\eqref{eq:Gi} and the coupling factor in Eq.~\eqref{eq:Def_Aa} are now
\[N_{a}=\frac{7}{27}\quad,\qquad N_{b}=-\frac{2}{27} \quad,\qquad N_g=4 g_s^6\,\,,\] 
where $g_s$ is the strong coupling constant. The internal masses are
\[m_{2}=m_{4}=m_{\tilde{g}}\quad,\qquad m_{1}=m_{\tilde{q}_{i}}\quad,\qquad m_{3}=m_{\,\tilde{t}_{j}}\,\,,\]
where $i$ and $j$ are the sfermion indices of the squark and the stop,
respectively. To make the dependence on the sfermion indices explicit
we use the notation
\[G_{q}^{a}\to G_{q,ij}^{a}\quad,\qquad A_{q}^{a}\to A_{q,ij}^{a}.\]
Ultimately, we must sum over the sfermion indices $i$ and $j$.
However, since both the internal masses and the couplings may depend
on the indices $i$ and $j$, we postpone that sum until the end of the
discussion. Due to $m_{2}=m_{4}$, we have the relation
\[A_{q,ij}^{+-++}=A_{q,ij}^{-+++},\]
as can be seen from the expressions for $D^a$ in Section~\ref{sec:da}. 

Using the unitarity of the squark mixing matrices of Eq.~\eqref{eq:squarkmix},
we can simplify sums and differences of the coupling products: 
\begin{align*}
G_{q,ij}^{+-++}+G_{q,ij}^{-+++} & =-2(N_{a}+N_{b})Re(U_{j,+}^{\tilde{t}}U_{j,-}^{\tilde{t}*}e^{i\phi})\equiv(N_{a}+N_{b})G_{j}^{(1)}\,\,,\\
G_{q,ij}^{-++-}+G_{q,ij}^{-+-+} & =(N_{a}+N_{b})\,\,,\\
G_{q,ij}^{-++-}-G_{q,ij}^{-+-+} & =(N_{a}-N_{b})(|U_{i,+}^{\tilde{q}}|^{2}-|U_{i,-}^{\tilde{q}}|^{2})(|U_{j,+}^{\tilde{t}}|^{2}-|U_{j,-}^{\tilde{t}}|^{2})\equiv(N_{a}-N_{b})G_{ij}^{(2)}
\end{align*}

Thus, we write for $A_{q}$ of Eq.~\eqref{eq:Hadron_AFBq} (again we added the sfermion indices $ij$) 
\begin{equation}
A_{q,ij}=Re\left\{G_{j}^{(1)}A_{q,ij}^{(1)}+G_{ij}^{(2)}A_{q,ij}^{(2)}+A_{q,ij}^{(3)}\right\}\label{eq:AFBqSQCD}
\end{equation}
with 
\begin{eqnarray}
A_{q,ij}^{(1)} & =&(N_{a}+N_{b})A_{q,ij}^{-+++} \nonumber \\
A_{q,ij}^{(2)} & =&(N_{a}-N_{b})\tfrac{1}{2}(A_{q,ij}^{-++-}-A_{q,ij}^{-+-+})
\nonumber \\
A_{q,ij}^{(3)} &
=&(N_{a}+N_{b})\tfrac{1}{2}(A_{q,ij}^{-++-}+A_{q,ij}^{-+-+}) \label{eq:AkqijSQCD}
\end{eqnarray}
and 
\begin{eqnarray}
G_{j}^{(1)} & =& -2Re(U_{j,+}^{\tilde{t}}U_{j,-}^{\tilde{t}*}e^{i\phi}) \nonumber \\
G_{q,ij}^{(2)} & =& (|U_{i,+}^{\tilde{q}}|^{2}-|U_{i,-}^{\tilde{q}}|^{2})(|U_{j,+}^{\tilde{t}}|^{2}-|U_{j,-}^{\tilde{t}}|^{2})\,.\label{eq:G1ijG2ij}
\end{eqnarray}
Using the unitarity of the stop mixing matrix $U^{\tilde{t}}$, we find 
\begin{equation}
G_{1}^{(1)}=-G_{2}^{(1)}\quad,\qquad G_{q,i1}^{(2)}=-G_{q,i2}^{(2)}\quad,\qquad G_{q,1j}^{(2)}=-G_{q,2j}^{(2)}\,\,.
\end{equation}
Note that $G_{j}^{(1)}$ and $G_{q,ij}^{(2)}$ are always real, and one
therefore only needs to consider the real part of the functions $A_{q,ij}^{a}$.
Furthermore, the coupling product $G_{j}^{(1)}$
does not depend on the initial-state quark flavor $q$.
If one assumes that the mixing matrix of the
light flavor squarks $\tilde{q}$ is diagonal, as in almost
all of the considered parameter space, $G_{q,ij}^{(2)}$
reduces to
\begin{equation}
 G_{q,ij}^{(2)}\,=\, (-1)^i\cdot (|U_{j,+}^{\tilde{t}}|^{2}-|U_{j,-}^{\tilde{t}}|^{2})
                 \equiv G_{ij}^{(2)}
\end{equation}
and does not depend on the initial-state quark flavor $q$ either.
We can thus perform the sum over $i$, $j$ and $q$ and obtain 
\begin{equation}
A_{\text{FB,SQCD}}^{t\bar t}=G_{1}^{(1)}A^{(1)}+G_{11}^{(2)}A^{(2)}+A^{(3)}
\end{equation}
with 
\begin{align}
A^{(1)} & =\sum_{q=u,d,s,c}Re[A_{q,11}^{(1)}-A_{q,12}^{(1)}+A_{q,21}^{(1)}-A_{q,22}^{(1)}]\,\,,\nonumber \\
A^{(2)} & =\sum_{q=u,d,s,c}Re[A_{q,11}^{(2)}-A_{q,12}^{(2)}-A_{q,21}^{(2)}+A_{q,22}^{(2)}]\,\,,\nonumber \\
A^{(3)} & =\sum_{q=u,d,s,c}Re[A_{q,11}^{(3)}+A_{q,12}^{(3)}+A_{q,21}^{(3)}+A_{q,22}^{(3)}]\,\,.\label{eq:loopfunc}
\end{align}
Note that $A^{(1)}$ vanishes for degenerate stop masses and $A^{(2)}$
vanishes if the stop or squark masses are degenerate. Upper and lower
limits on the SUSY QCD one-loop contribution to $A_{\text{FB}}^{t\bar
t}$ can now be obtained for given values of the squark masses by
exploiting the fact that the limits of $G_{1}^{(1)}$ and
$G_{11}^{(2)}$ are
\[-1\leq G_{1}^{(1)}\leq 1 \; , \; -1\leq G_{11}^{(2)}\leq 1\]
and thus
\begin{equation}
A^{(3)}-|A^{(1)}|-|A^{(2)}|\leq A_{\text{FB,SQCD}}^{t\bar t} \leq A^{(3)}+|A^{(1)}|+|A^{(2)}|\,\,.
\label{eq:sqcdlimits}
\end{equation}
%
\subsection{SUSY EW One-loop Contributions
\label{sec:sew}}
\begin{figure}[h]
\includegraphics{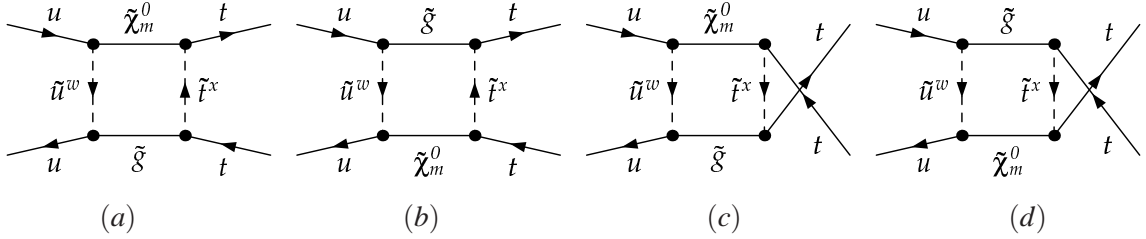}

$(a)$\hspace*{3.5cm}$(b)$\hspace*{3.5cm}$(c)$\hspace*{3.5cm}$(d)$

\caption{Feynman diagrams of the SUSY EW one-loop contribution to the forward-backward asymmetry $A_{\text{FB}}^{t\bar t}$.
  \label{fig:Diags_EW}}
\end{figure}

%
The SUSY EW one-loop contributions to the forward-backward asymmetry
$A_{\text{FB}}^{t\bar t}$ of Eq.~\eqref{eq:afbtt} consists of four
diagrams, two direct boxes and two crossed boxes, shown in
Fig.~\ref{fig:Diags_EW}.
We again start with the generic expression for $A_{\text{FB}}^{t\bar
t}$ in terms of loop functions and coupling products as given in
Eq.~\eqref{eq:Hadron_AFBq}. The color factors in Eq.~\eqref{eq:Gi} and
the coupling factor in Eq.~\eqref{eq:Def_Aa} are now
\[N_{a}=N_b=\frac{2}{9} \quad,\qquad N_g=4 g_s^4 e^2\,\,,\] 
where $e$ is the electromagnetic coupling. The internal masses are 
\begin{eqnarray*}
m_{1}\,\,=\,\, m_{\tilde{q}_{i}}\,,\,\,\, m_{3}=m_{\,\tilde{t}_{j}},\\
{\rm Fig.}\,\ref{fig:Diags_EW}(a,c)\!:\quad m_{2}\,\,=\,\,
m_{\tilde{\chi}_{k}}\,,\,\,\, m_{4}\,\,=\,\, m_{\tilde{g}} & \quad{\rm
or}\quad & {\rm Fig.}\,\ref{fig:Diags_EW}(b,d)\!:\,\, m_{2}\,\,=\,\,
m_{\tilde{g}}\,,\,\,\, m_{4}\,\,=\,\,
m_{\tilde{\chi}_{k}}
\end{eqnarray*} 
where $i$ and $j$ are the sfermion indices of the squark and the stop,
respectively, and $k$ is the neutralino index. To make the dependence
on the sfermion and neutralino indices explicit, we use the notation
\begin{eqnarray*}
G_{q}^{a}\to G_{q,kij}^{a}\quad,\qquad A_{q}^{a}\to
A_{q,kij}^{a} \, .
\label{eq:Def_Aew}
\end{eqnarray*}
Ultimately, we must sum over the indices $i$, $j$ and $k$. This are $16$
combinations for each diagram.  By examining the amplitudes for the
  individual diagrams one notices that the diagrams (a) and (c) in
  Fig.~\ref{fig:Diags_EW} are the complex conjugates of the diagrams (b) and
  (d), respectively. Thus, the $A^a_{q,ij}$ are real and  only the real
  parts of the coupling functions $G^{a}_q$ of Eq.~\eqref{eq:Gi} contribute. The contribution of the SUSY EW
  one-loop corrections to $A_{\text{FB}}^{t\bar t}$ of Eq.~\eqref{eq:Hadron_AFBq} then
  reads
\begin{eqnarray}
A_{FB,SEW}^{t\bar t} & = & \sum_q A_q \nonumber \\
&=& \sum_q \sum_{k=1}^{4}\sum_{i,j=1}^{2}\left[ \quad 
Re\left\{ G_{q,kij}^{\,+-++}\right\} A_{q,kij}^{+-++}+
Re\left\{ G_{q,kij}^{\,-+++}\right\} A_{q,kij}^{-+++}\right.\\
 &  &\hspace*{1.5cm}\left. {}+Re\left\{ G_{q,kij}^{\,-++-}\right\} A_{q,kij}^{-++-}
+Re\left\{ G_{q,kij}^{\,-+-+}\right\} A_{q,kij}^{-+-+}\right]
\label{eq:Asewq}
\end{eqnarray}
In case of SUSY EW one-loop corrections, it is quite complicated to
 analytically derive bounds on the coupling factors $G_{q,kij}^{a}$
because of the complicated structure of the neutralino-squark-quark
couplings (see Eq.~\eqref{eq:def_hat_g_Neut}). It is even harder then to
find reasonable bounds on $A_{FB,SEW}^{t\bar t}$ as we did in the SUSY
QCD case. The box diagrams have 16 different squark und neutralino
mass combinations where always some cancellation occurs due to the
unitarity of the mixing matrices. Therefore, we performed a MSSM
parameter scan to extract bounds on $A_{FB,SEW}^{t\bar t}$ as
described in Section~\ref{sec:results:mssm}.
%
\section{Numerical Results}\label{sec:results}
%
For the numerical evaluation of the forward-backward asymmetry
$A_{\text{FB}}^{t\bar t}$ of Eq.~\eqref{eq:afbtt}, we use the LO PDF set
CTEQ6L1~\cite{Pumplin:2002vw} with the renormalization ($\mu_R$) and
factorization ($\mu_F$) scales chosen to be equal to the top quark
mass, $\mu_R =\mu_F =m_t$.  The SM input parameters are
$m_{t}=173.2$~GeV, $\alpha=1/137.036$, $M_W=80.36$~GeV,
$M_Z=91.187$~GeV, and $\cos\theta_W=M_W/M_Z$. We assume one-loop
running of the strong coupling constant with $\alpha_s(M_Z)=
0.130$, so that $\alpha_s(m_t)=0.118$, which is consistent
with our choice of PDFs.  To evaluate the coefficients of the tensor
integrals $D_{i,ij}$ of Eq.~\eqref{eq:Definition_of_Da}, the LoopTools
library~\cite{Hahn:1998yk} has been used.

The study of the dependence of $A_\text{FB}^{t\bar t}$ on the MSSM
input parameters is simplified by the fact that it is not sensitive to
all parameters of the complex MFV-MSSM. First of all, the
forward-backward asymmetry can only come from diagrams with up or down
quarks in the initial state. For strange, charm and bottom quarks, the
PDFs are the same as those of the corresponding anti-quarks so that
the difference of PDFs in Eq.~\eqref{eq:hadron_AFB_a} is zero. In the
MFV-MSSM $A_\text{FB}^{t\bar t}$ is therefore insensitive to
parameters that only affect the masses and couplings of strange, charm
and bottom-squarks. Furthermore, the trilinear couplings of up and
down-squarks only enter 
through the squark mass matrices, where they
are suppressed by the small up and down-quark Yukawa couplings. Thus,
only the following set of MSSM input parameters are relevant to our
study:
\begin{itemize}
\item $\tan\beta$
\item $\mu$
\item the pseudoscalar Higgs mass $m_A$
\item the gaugino masses $M_1$, $M_2$ and $M_3$
\item the top-squark trilinear coupling $A_t$
\item the soft masses $m_{\tilde q_{L1}}$ and $m_{\tilde q_{L3}}$ of the
  left-handed first and third generation squarks
\item the soft masses $m_{\tilde u_R}$, $m_{\tilde d_R}$ and $m_{\tilde t_R}$ of
  the right-handed up, down and top squarks
\end{itemize}
Of these parameters, $\mu$, $M_1$, $M_2$, $M_3$ and $A_t$ can be
complex, but one of these phases can be rotated away. We rotate
the phase of $M_2$ away and study the dependence on the
remaining complex phases and the absolute values of the 
above MSSM parameters independently.

These input parameters are constrained by direct SUSY searches at LEP,
Tevatron and LHC, and indirectly by low-energy precision observables.
A review of results from the search for signals of low-energy SUSY at
LEP and the Tevatron as well as in precision observables can be found,
e.~g., in Ref.~\cite{Nakamura:2010zzi}. Most recently stringent
exclusion limits on squark and gluino masses within the CMSSM  and
Simplified Models have been obtained at the LHC by the
CMS~\cite{Chatrchyan:2011zy,Chatrchyan:2011ek,Khachatryan:2011tk} and
ATLAS~\cite{Aad:2011ib,ATLAS:2011ad,Aad:2011hh,daCosta:2011qk}
collaborations.  In Section~\ref{sec:results:mssm} we provide general
upper and lower bounds on $A_\text{FB}^{t\bar t}$ within the complex
MFV-MSSM by performing a comprehensive scan over a wide range of
values for these input parameters without any additional assumptions.
In particular, these general bounds do not rely, for instance, on
specific SUSY breaking scenarios or an artificially reduced parameter
space such as the CMSSM.  The results are presented in such a way that
the effect of a change in the sparticle mass limits on the upper and
lower bounds on $A_{\text{FB}}^{t\bar t}$ can be estimated. In
Section~\ref{sec:results:mssm} this is done for the current LHC squark
and gluino mass limits.

In the next sections we first discuss the main characteristics of
the MSSM one-loop corrections to $A_{\text{FB}}^{t\bar t}$ and then
present results of a comprehensive parameter scan.  

\subsection{SUSY QCD and SUSY EW Contributions to the Loop Functions \label{sec:results:loop}}

The main characteristics of the MSSM one-loop corrections to, and of
bounds on, $A_{\text{FB}}^{t\bar t}$ can be determined by studying
the SUSY QCD loop functions $A_{q,ij}^{(k)}$ of
Eq.~\eqref{eq:AkqijSQCD} and SUSY EW loop functions $A_{q,ijk}^{a}$ of
Eq.~\eqref{eq:Asewq}.  Here we will only present results for the
contribution of the $u\bar u$-initiated $t\bar t$ production process,
since the $d\bar{d}$ production channel is much smaller due to the
smaller PDF (about a factor of eight smaller as discussed in
Section~\ref{sec:results:mssm}).

\begin{figure}[hbt]
\hspace*{-4mm}\includegraphics[scale=0.4]{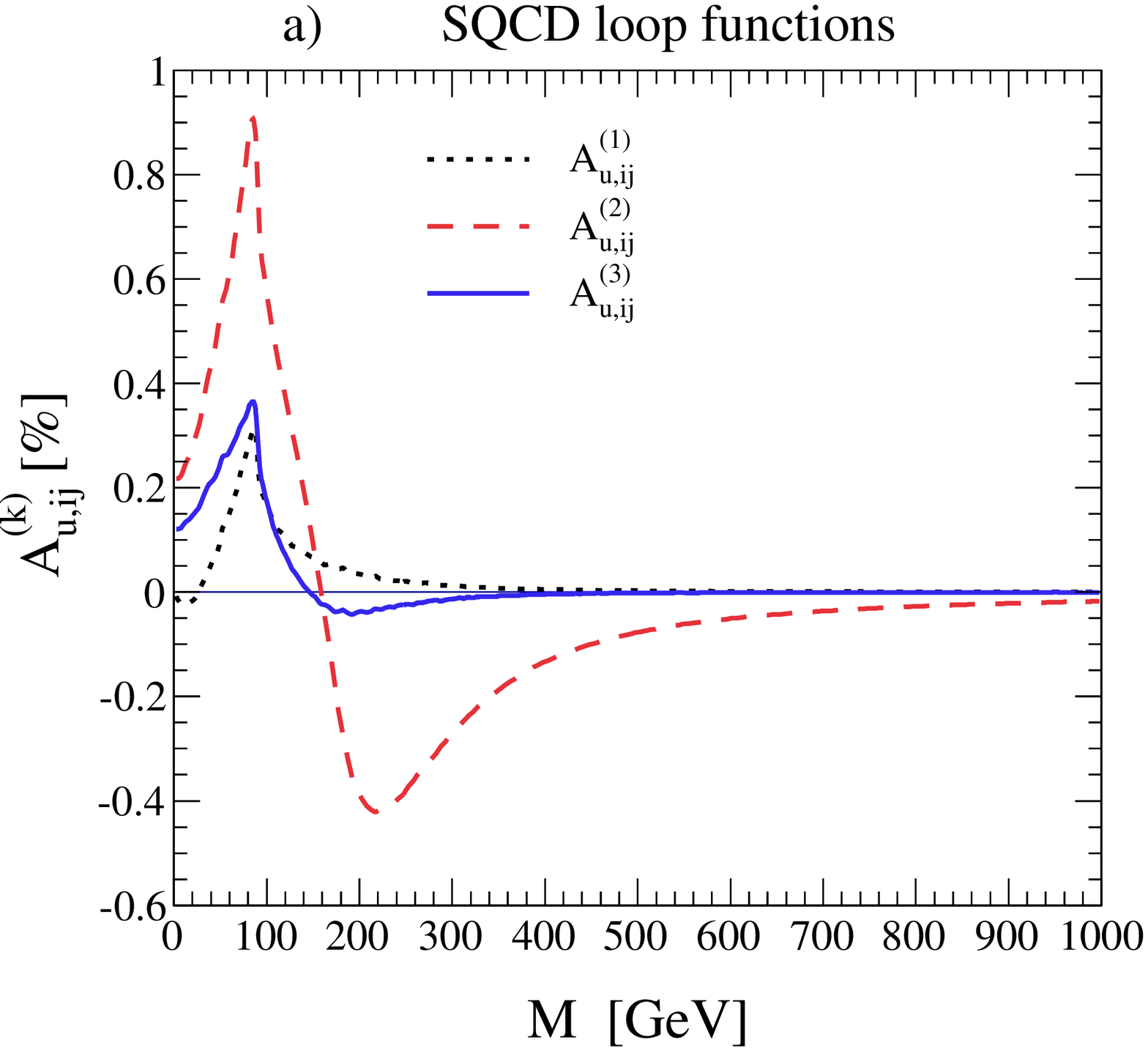}
\hspace*{-2mm}\includegraphics[scale=0.4]{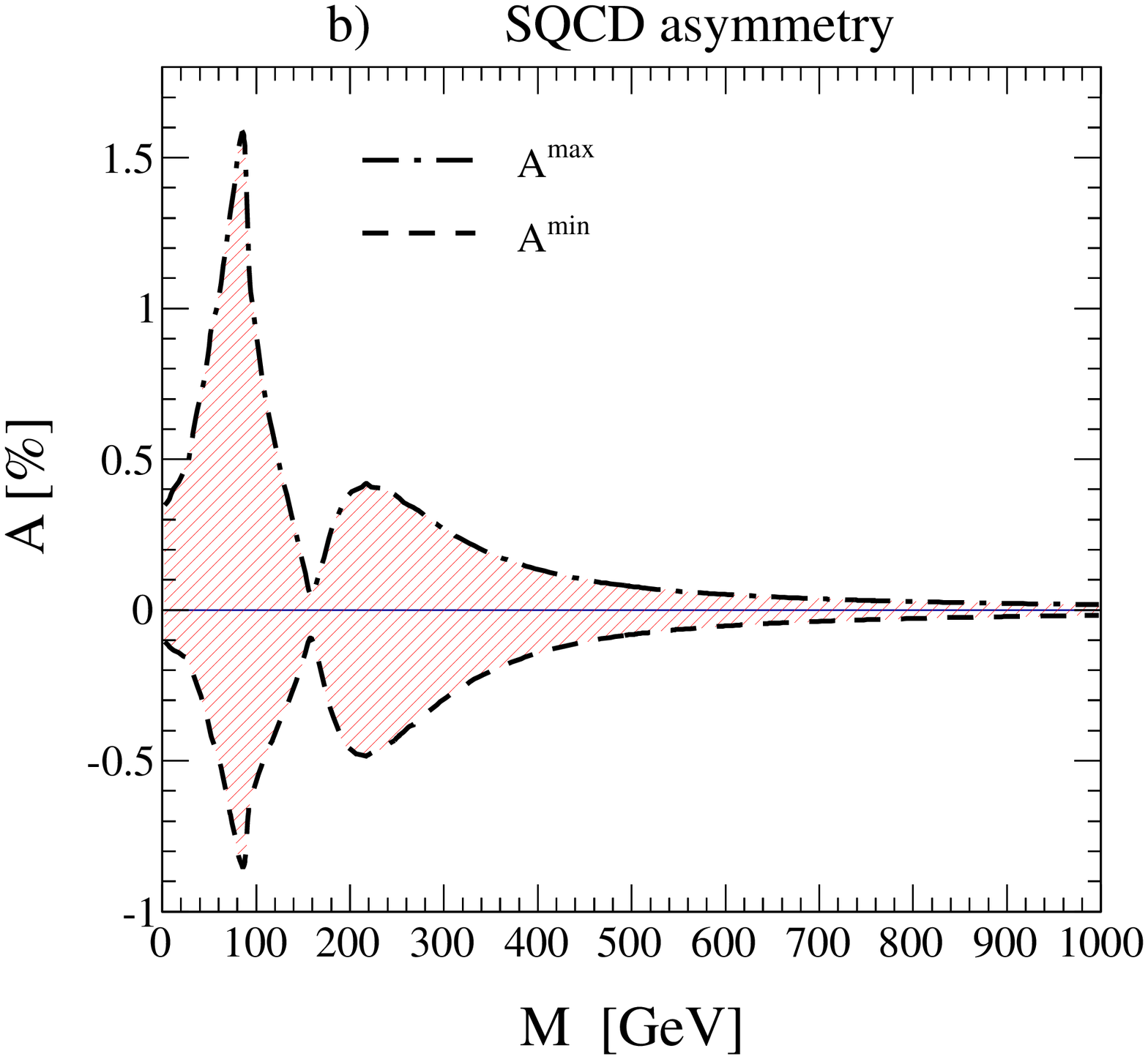}\hspace*{-5mm}\mbox{}\\[-8pt]
\caption{(a) SUSY QCD normalized hadronic loop functions $A^{(k)}_{u,ij}$
  with $k=1,2,3$ as defined in Eq.~\eqref{eq:AkqijSQCD} and (b) 
  bounds $A_{min,max}$ on $A_{\text{FB,SQCD}}^{t\bar t}$
  of Eq.~\eqref{eq:sqcd_aminmax} for initial-state up-quarks. Shown is the
  dependence on a common mass $M$. The bounds are obtained assuming large 
  up-squark and top-squark mass splittings. No kinematic cuts have been applied.}
\label{fig:m_afbsqcd} 
\end{figure}

\begin{figure}[hbt]
\hspace*{-4mm}\includegraphics[scale=0.4]{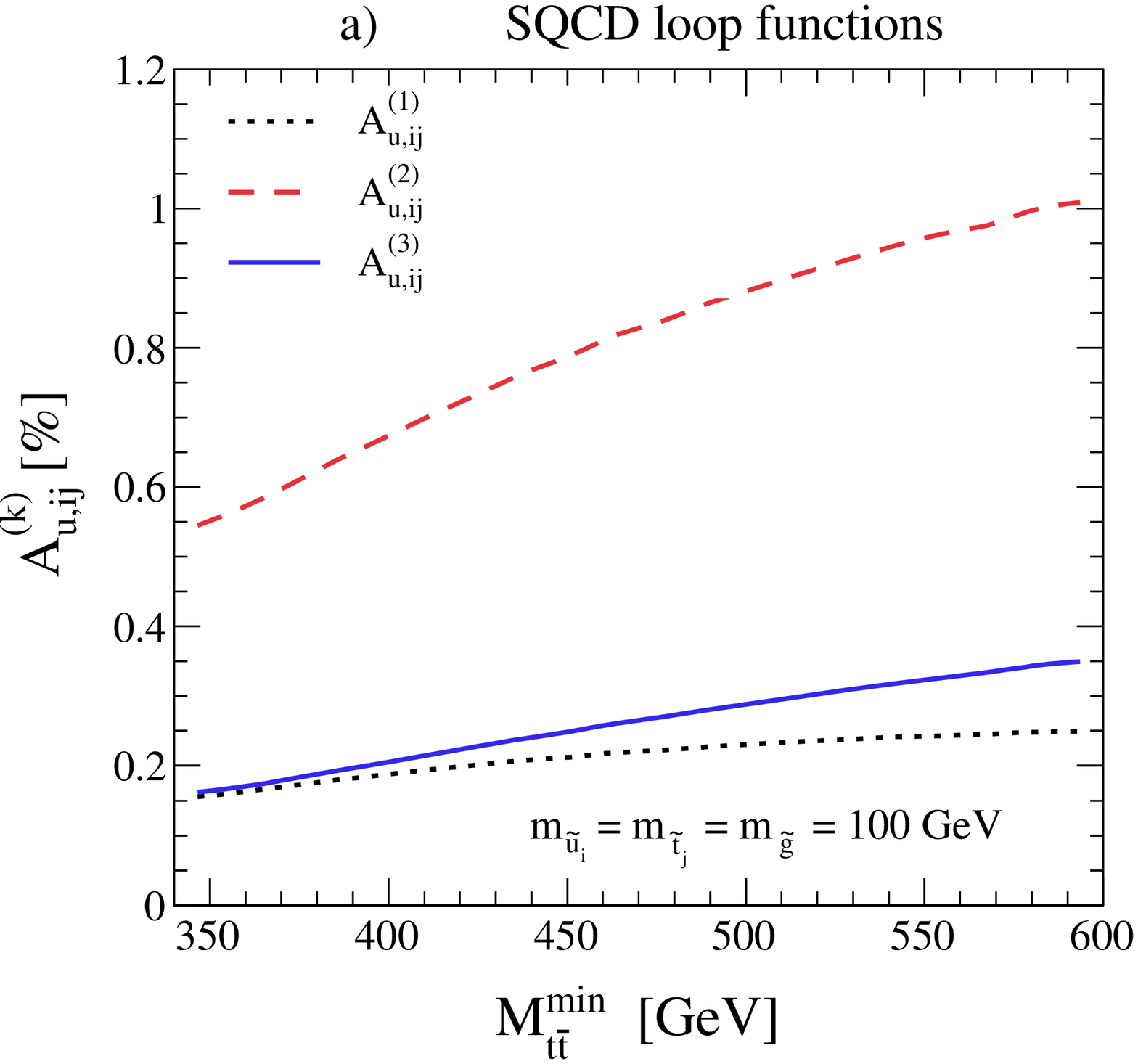}
\hspace*{-2mm}\includegraphics[scale=0.4]{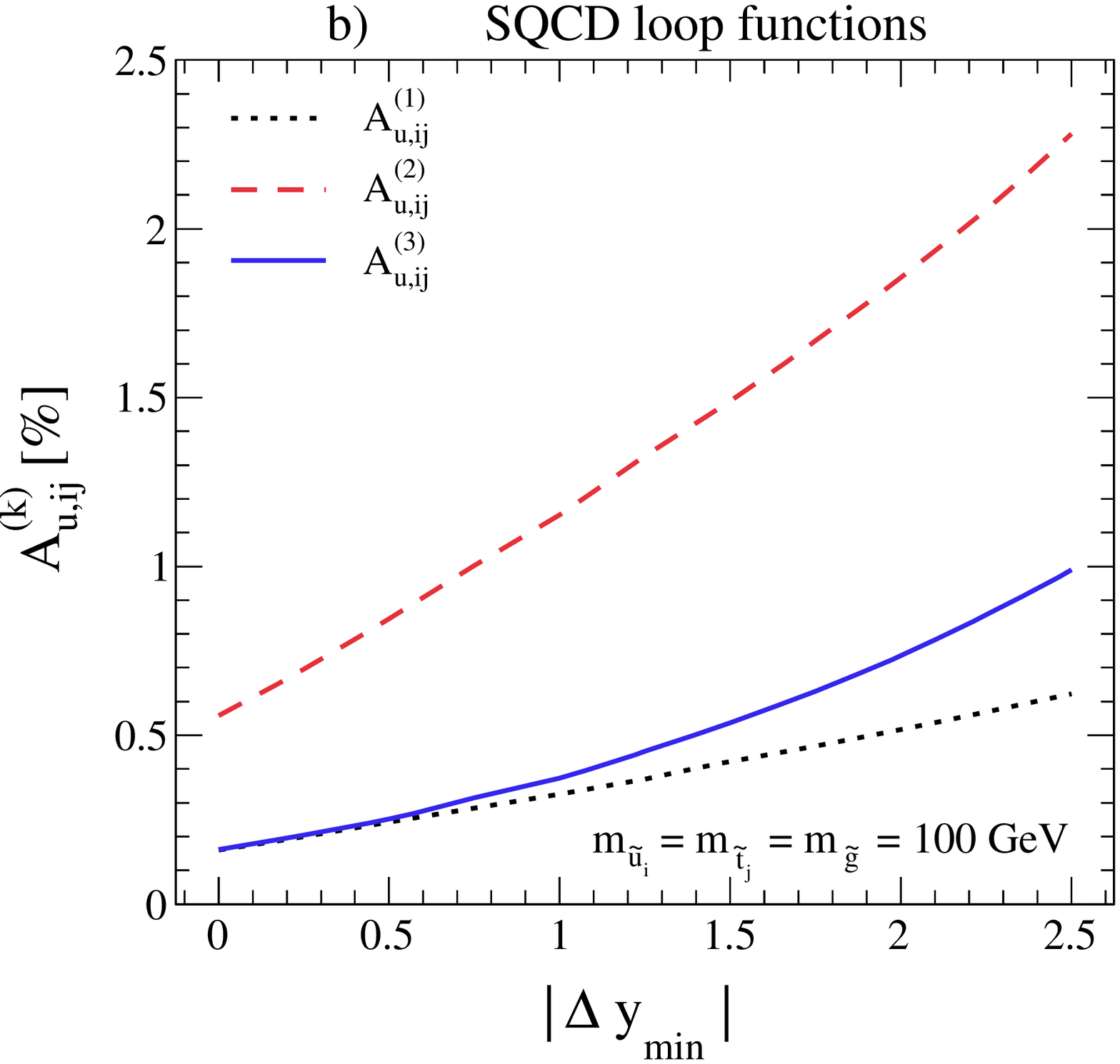}\hspace*{-5mm}\mbox{}\\[-8pt]
\caption{SUSY QCD normalized hadronic loop functions $A^{(k)}_{u,ij}$
with $k=1,2,3$ for initial-state up-quarks as defined in
Eq.~\eqref{eq:AkqijSQCD}, assuming all particles in the loop have a
common mass $M=100$~GeV. Shown is the dependence on the cut 
on (a)  the invariant $t\bar t$ mass $M_{t\bar t}$, $M_{t\bar
  t}>M^{min}_{t\bar t}$, and on (b)
 the rapidity
difference $|\Delta y| = |y_{t} - y_{\bar{t}}|$, $|\Delta y| > \Delta
y_{min}$.}
\label{fig:m_afbsqcd_cuts} %
\end{figure}

In case of SUSY QCD contributions we showed in Section~\ref{sec:sqcd}
that bounds on $A_{FB,SQCD}^{t \bar t}$ of Eq.~\eqref{eq:AkqijSQCD}
can be derived in terms of the normalized hadronic loop functions
$A_{q,ij}^{(k)}, k=1,2,3$, where $q$ denotes the initial-state quark
flavor, $i$ the squark index $i=1,2$ with flavor $q$, and $j=1,2$
refers to the top squark index.  These loop functions only depend on
three general mass parameters, $m_1=m_{\tilde q_i}$, $m_2=m_{\tilde
g}$, and $m_3=m_{\tilde t_j}$, as indicated in
Figs.~\ref{fig:boxes},\ref{fig:Feynman_Diags_SQCD}. For the special
case of initial-state up-type quarks and mass degeneracy,
$m_1=m_2=m_3=M$, we show numerical results for the functions
$\smash{A_{u,ij}^{(k)}}$ ($k=1,2,3$) in
Figs.~\ref{fig:m_afbsqcd}(a)~and~\ref{fig:m_afbsqcd_cuts}.
As can be seen in Fig.~\ref{fig:m_afbsqcd}(a), the largest single
contribution to the loop functions comes from $A_{u,ij}^{(2)}$. This
feature persists when the masses are varied
independently. $A_{u,ij}^{(2)}$ reaches up to ${}+0.9\%$ for
$M\approx\unit{100}$~GeV and ${}-0.5\%$ for $M\approx\unit{200}$~GeV.
The other two loop functions can reach roughly ${}+0.4\%$ for
$M\approx\unit{100}$~GeV and only tiny negative values. For masses
$M\ge\unit{400}$~GeV all contributions are very small. The first
peak in Figs.~\ref{fig:m_afbsqcd}(a) and (b) at $M=86.6$~GeV is due to
a normal threshold~\cite{Landau:1959fi} when the condition $p_3^2 =
(m_2+m_3)^2$ or $p_4^2 = (m_3+m_4)^2$ is fulfilled, which here is the
case when $M=m_t/2$.  A resonance in the partonic $t\bar t$ cross
section occurs when the gluino pair in the box diagrams can become
on-shell, thus $m_{\tilde{g}}> m_t$. This resonance manifests itself
as dips in the hadronic functions $A_{u,ij}^{(k)}$ at $M\approx
200$~GeV. When the resonance occurs inside the numerically important
$t\bar{t}$ invariant mass region of $\sqrt{\hat{s}}= 350$ to
$500$~GeV, it leads to a correspondingly large negative asymmetry.
For larger SUSY particle masses $M$, the resonance region is shifted to
larger values of $\sqrt{\hat{s}}$ and therefore outside the dominant
region of the $t\bar{t}$ cross section.  From these results an
estimate of the bounds $A_{min,max}$ on $A_{\text{FB,SQCD}}^{t\bar t}$
can be obtained using
Eqs.~\eqref{eq:loopfunc}~and~\eqref{eq:sqcdlimits}. As noted earlier,
$A^{(1)}$ vanishes for degenerate stop masses and $A^{(2)}$ vanishes
if the stop or squark masses are degenerate.  In general, we found that
the SUSY QCD one-loop corrections to the forward-backward asymmetry
increase if the up-squark mass splitting or, in particular, the
top-squark mass splitting are increased.  Thus, the largest asymmetry
is obtained if $\tilde u_2$ and $\tilde t_2$ are decoupled, which
results in vanishing functions $A^{(k)}_{u,12}$, $A^{(k)}_{u,21}$, and
$A^{(k)}_{22}$. In this scenario the bounds of
Eq.~\eqref{eq:sqcdlimits} read:
\begin{eqnarray}
A_{min}&=& A^{(3)}_{u,11}-|A^{(1)}_{u,11}|-|A^{(2)}_{u,11}| \nonumber \\
A_{max}&=& A^{(3)}_{u,11}+|A^{(1)}_{u,11}|+|A^{(2)}_{u,11}|\,\,.\label{eq:sqcd_aminmax}
\end{eqnarray}
Figure~\ref{fig:m_afbsqcd}(b) shows the dependence of these bounds
$A_{min,max}$ on $M=m_{\tilde t_1}=m_{\tilde u_1}=m_{\tilde g}$. The
total asymmetry can reach values from ${}-0.9\%$ up to ${}+1.6\%$ for
this configuration when no kinematic cuts have been applied.

Figures.~\ref{fig:m_afbsqcd_cuts}(a) and (b) show the loop functions
$A_{u,ij}^{(k)}$ in dependence of a  cut on $M_{t\bar{t}}$,
$M_{t\bar t}^{min}$, and on $|\Delta y|$, $\Delta y_{min}$,
respectively.  From $M^{min}_{t\bar{t}}=350$~GeV to
$M^{min}_{t\bar{t}}=600$~GeV, the functions $A_{u,ij}^{(2)}$ and
$A_{u,ij}^{(3)}$ roughly double in size while $A_{u,ij}^{(1)}$
increases by about $50\%$.  If one increases $M^{min}_{t\bar{t}}$ from
$350$~GeV to $450$~GeV, the loop functions increase by a factor of
$1.35$ for $A_{u,ij}^{(1)}$, $1.43$ for $A_{u,ij}^{(1)}$ and $1.52$
for $A_{u,ij}^{(3)}$.  Thus, the bounds on the forward-backward
asymmetry for the SUSY QCD contributions will roughly change by a
factor of $1.4$ if one applies a $M^{min}_{t\bar{t}}=450$~GeV cut
compared to the case without such a cut.  Applying a cut on $|\Delta
y|$, as shown in Fig.~\ref{fig:m_afbsqcd_cuts}(b), also increases the
loop functions $A_{u,ij}^{(k=1,2,3)}$.  For instance, when changing
$\Delta y_{min}$ from zero to one, $A_{u,ij}^{(1,2)}$ increase by a
factor of~$\,2.00$ and $A_{u,ij}^{(3)}$ by a factor of $2.25$.

\begin{figure}[hbt]
\includegraphics[scale=0.40]{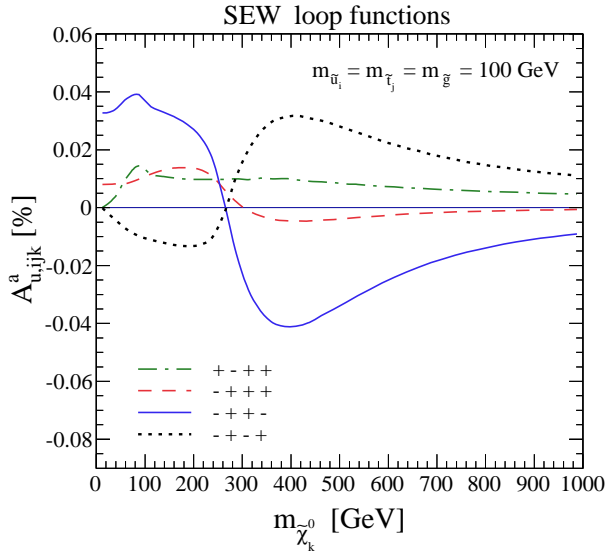}\hspace*{-5mm}\mbox{}\\[-8pt]
\caption{SUSY EW normalized hadronic loop functions $A^{a}_{u,ijk}$
with $a \in \{+-++,-+++,-++-,-+-+\}$ for initial-state up-quarks as
defined in Eq.~\eqref{eq:Def_Aew} in the case where the up-squark, top
squark and the gluino have a common mass $M =
m_{\tilde{g}}=m_{\tilde{t}_j}=m_{\tilde{u}_i}=100$~GeV. Shown is the
dependence on the neutralino mass $m_{\chi^0_k}$. No kinematic cuts
have been applied.}
\label{fig:m_afbsew} %
\end{figure}

We now investigate the loop functions of the SUSY EW one-loop
contributions to the forward-backward asymmetry as described by
$A_{FB,SEW }^{t\bar t}$ of Eq.~\eqref{eq:Asewq}.  In this case, one
has four different masses in the loop and four different loop
functions $A^{a}_{u,ijk}$ with $a \in \{+-++,-+++,-++-,-+-+\}$. If one
plots these loop functions in dependence of a common SUSY mass $M$,
one can produce a similar plot as in Fig.~\ref{fig:m_afbsqcd}, just
rescaled by the smaller coupling factor and different color factors.
To illustrate the effect of the neutralino mass we therefore show in
Fig.~\ref{fig:m_afbsew} the loop functions for the SUSY EW one-loop
contributions as defined in Eq.~\eqref{eq:Def_Aew} in dependence of
the neutralino mass, when assuming a common mass for the other
sparticles in the loop: $M
=m_{\tilde{g}}=m_{\tilde{t}_j}=m_{\tilde{u}_i}=100$~GeV.  The loop
functions $A^{-++-}_{u,ijk}$ and $A^{-+-+}_{u,ijk}$ give the largest
contributions. They can amount to about $\pm 0.04\%$ and only slowly
decrease with increasing neutralino mass, e.~g., they are still about
$\pm 0.02\%$ for relatively large neutralino masses
$m_{\tilde{\chi}^0_k} \approx 700$~GeV.

As stated in Section~\ref{sec:sew}, it is difficult to obtain an
analytic expression for bounds on $A_{FB,SEW}^{t\bar t}$ like it
was done in case of SUSY QCD-induced asymmetries. We refer therefore
to the next section where we perform a comprehensive scan over the
relevant MFV-MSSM parameter space to determine the largest
possible value of $A_{\text{FB}}^{t\bar t}$ induced by both SUSY QCD and SUSY
EW one-loop corrections.

%
\subsection{Bounds on $A_{\text{FB}}^{t\bar t}$ from a MSSM parameter scan}
\label{sec:results:mssm}

In phenomenological studies of the MSSM the large number of parameters
is a common problem. Even if all parameters of the soft MSSM
Lagrangian are assumed to be real and all flavor structures are
assumed to be proportional to the SM Yukawa matrices (minimal flavor
violation) we are left with 30 independent
parameters~\cite{Djouadi:2002ze}. Numerical discussions of observables
within the MSSM are therefore often limited to constrained scenarios,
where certain assumptions about the SUSY breaking mechanism are
imposed, or even to individual benchmark points which are deemed
``representative'' in some sense. Here we describe in detail how we
performed a comprehensive scan over the  relevant parameter space 
of the complex
MFV-MSSM to determine bounds on $A_{\text{FB}}^{t\bar t}$. 

In the case of $A_\text{FB}^{t\bar t}$, we are actually able to scan
the full  relevant parameter space of the complex MFV-MSSM. This is made
possible by the separation of $A_\text{FB}^{t\bar t}$ into loop
functions $A_{q}^a$ and products of coupling parameters $G_q^a$, which
is described in Section~\ref{sec:MSSM-Contributions-to_Afb}.
Calculating the loop functions of Eq.~\eqref{eq:hadron_AFB_a} requires a
numerical phase space integration and is therefore rather
time-consuming. However, the $A_{q}^a$ are relatively smooth functions
of only four variables (namely, the internal masses
$m_1,\ldots,m_4$). The loop functions may therefore be calculated on a
four-dimensional ``mass-grid'' and linear interpolation can be used to
obtain $A_{q}^a$ for other mass values within the grid range. The
remaining computational cost of calculating masses and coupling
constants for a given set of MSSM parameters and interpolating the
loop functions is extremely small, so that sampling up to one billion
(!) MSSM parameter points is perfectly doable on a single core
computer.

Let us briefly discuss the details of the grid interpolation and the
parameter scan. Grid data for the loop functions were generated for
masses between $0$ and $2$~TeV. For masses below $500$~GeV the grid
spacing was $20$~GeV. If one of the masses exceeds $500$~GeV the grid
spacing was increased to $50$~GeV in that direction.  If a mass
exceeds $1$~TeV the grid spacing was increased again to $100$~GeV.  At
each grid point the integrals in Eq.~\eqref{eq:hadron_AFB_a} were
calculated with the VEGAS algorithm.  Specifically, we use the
OmniComp-Dvegas package~\cite{DVEGAS}, which facilitates parallelized
adaptive Monte Carlo integration and was developed in the context of
~\cite{Kauer:2001sp,Kauer:2002sn}. A lower cut of $450$~GeV on the
$t\bar t$ invariant mass was applied throughout. Thus, all the
results in this section are for the `large $m_{t\bar t}$' bin.
Separate integrations of $A_{q}^a$ were done for each value of the
superscript $a$ (three for the SUSY QCD contributions and four for the
SUSY EW contributions), but the values for different quark flavors $q$
were determined with the same simulation.  Adaptation was driven by
the $u$-quark flavor, which always produces the largest value. The
relative accuracy of the numerical integration was required to be
below 1\%.

The multivariate linear interpolation was done by successively using
one-dimensional linear interpolation in each of the variables. If,
during the parameter scan, a certain mass exceeds $2$~TeV the loop
functions where it enters are assumed to be zero. In doing this, we
neglect loop functions of the order of one permille. The discrepancies
between the exact and the interpolated values of the loop functions
are of the same order.

As discussed earlier, the parameter scan is simplified by the fact
that $A_\text{FB}^{t\bar t}$ is not sensitive to all parameters of the
complex MFV-MSSM. Our choice of  relevant MSSM input parameters
is listed in the beginning of Section~\ref{sec:results}. The value of
$\tan\beta$ was varied between $1$ and $50$ and all parameters with
mass dimension one were varied between 0 and $3$~TeV. The complex
phases of $\mu$, $M_1$, $M_3$ and $A_t$ were varied between 0 and
$2\pi$.

For the actual scan we used an adaptive method along the lines
of Ref.~\cite{Brein:2004kh}.  The basic idea is the following: Instead of
sampling all parameters with a uniform (or otherwise fixed) random
distribution, one defines an importance function which quantifies the
importance of a given set of parameters. Since we are interested in
MSSM parameter points with large effects in $A_\text{FB}^{t\bar t}$ we
used $|A_\text{FB}^{t\bar t}|$ as importance function. We then used
VEGAS to compute the integral of the importance function over all the
scan parameters. Adaptation guarantees that the ``important'' regions
of the parameter space are sampled with a higher density. The
OmniComp-Dvegas package~\cite{DVEGAS} was used for the parameter scan
too.  A total of $8\cdot 10^9$ parameter points were
sampled. Adaptation was done with 22 iterations.

With the data from our scan, we can now show results for upper and
lower bounds on $A_\text{FB}^{t\bar t}$ as a function of any 
relevant MSSM input parameter. To do this, we simply bin the sample
points with respect to that input parameter and determine the maximal
and minimal value of $A_\text{FB}^{t\bar t}$ in each bin.

\begin{figure}[htb]
\hspace*{-6mm}\includegraphics[scale=0.41]{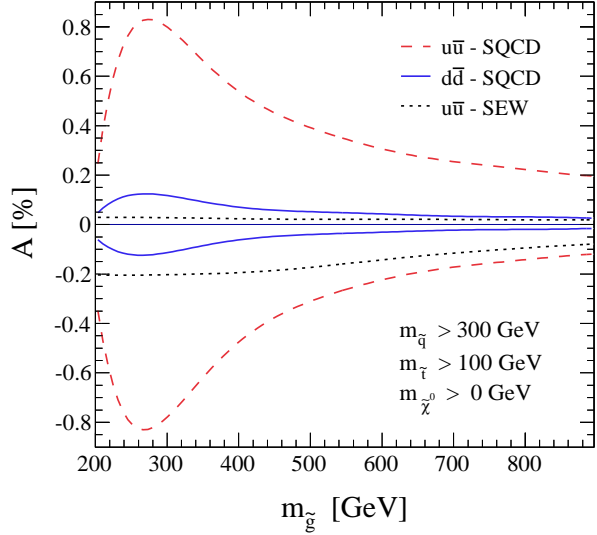}
\hspace*{-5mm}\mbox{}\\[-8pt]
\caption{The gluino mass dependence of the lower and upper bound on
$A_{\text{FB}}^{t\bar t}$, $|A_{\text{FB}}^{t\bar t}| \le A$, due to SUSY QCD and
SUSY EW one-loop contributions, with $M_{t\bar{t}}> 450$~GeV.
Separately shown are the bounds when only including the $u\bar u$ and
$d\bar d$-initiated $t\bar t$ production channels.}
\label{fig:afbmax_gluino} %
\end{figure}

Fig.~\ref{fig:afbmax_gluino} shows the upper and lower bounds on
$A_\text{FB}^{t\bar t}$ as a function of the gluino mass.  Shown
separately are the contributions from the $u\bar u$ and $d\bar
d$-initiated $t\bar t$ production channels as well as from SUSY QCD
and SUSY EW one-loop corrections, assuming $M_{t\bar{t}}>450$~GeV.

The bounds shown in Fig.~\ref{fig:afbmax_gluino} have been obtained by assuming
a lower limit on the top squark masses of $m_{\tilde{t}_{1,2}} > 100$~GeV and a
mass limit for the other squark flavor masses of $m_{\tilde{q}_{1,2}} >
300$~GeV. For the neutralinos no mass limit was imposed. As discussed
earlier, the $u\bar{u}$-channel generates the largest contribution to
$A_{\text{FB}}^{t\bar t}$ due to the large PDF and is roughly a factor of eight larger
than contribution from the $d\bar d$-channel.  The SUSY QCD bounds peak around
$m_{\tilde{g}} \approx 270$~GeV, when the contribution of the resonant gluino
pair in the box diagrams coincides with the maximum of the LO
  $M_{t\bar t}$ distribution. The absolute values of the upper and lower bounds
are practically identical because for the applied squark limits the SUSY QCD
loop function $A_{q,ij}^{(2)}$ is dominant (see Fig.~\ref{fig:m_afbsqcd}), so
that the SUSY QCD bound of Eq.~\eqref{eq:sqcdlimits} is well approximated by
$-|A^{(2)}|\leq A_{\text{FB,SQCD}}^{t\bar t} \leq |A^{(2)}|$.

As shown in Fig.~\ref{fig:afbmax_gluino}, the bounds on the SUSY EW
one-loop corrections to $A_{\text{FB}}^{t\bar t}$ are much
smaller than the SUSY QCD one-loop corrections, since they are
suppressed by the smaller electroweak coupling.  Furthermore, the
bounds are not as symmetric as in the SUSY QCD case. In general we
found the absolute value of the lower bound to be larger than the
upper bound. The SUSY EW one-loop corrections to the
$d\bar{d}$-channel are not shown in Fig.~\ref{fig:afbmax_gluino} as
they are basically zero, being suppressed by both the $d$-quark PDF
and the electroweak coupling. Again, larger SUSY EW loop-induced
asymmetries can be obtained when relaxing the constraints on the MSSM
parameters. For instance, for sparticle masses below $50$~GeV 
one can obtain $A_{FB,SEW}^{t\bar t}=-0.4 \%$.

\begin{figure}[htb]
\hspace*{-6mm}\includegraphics[scale=0.41]{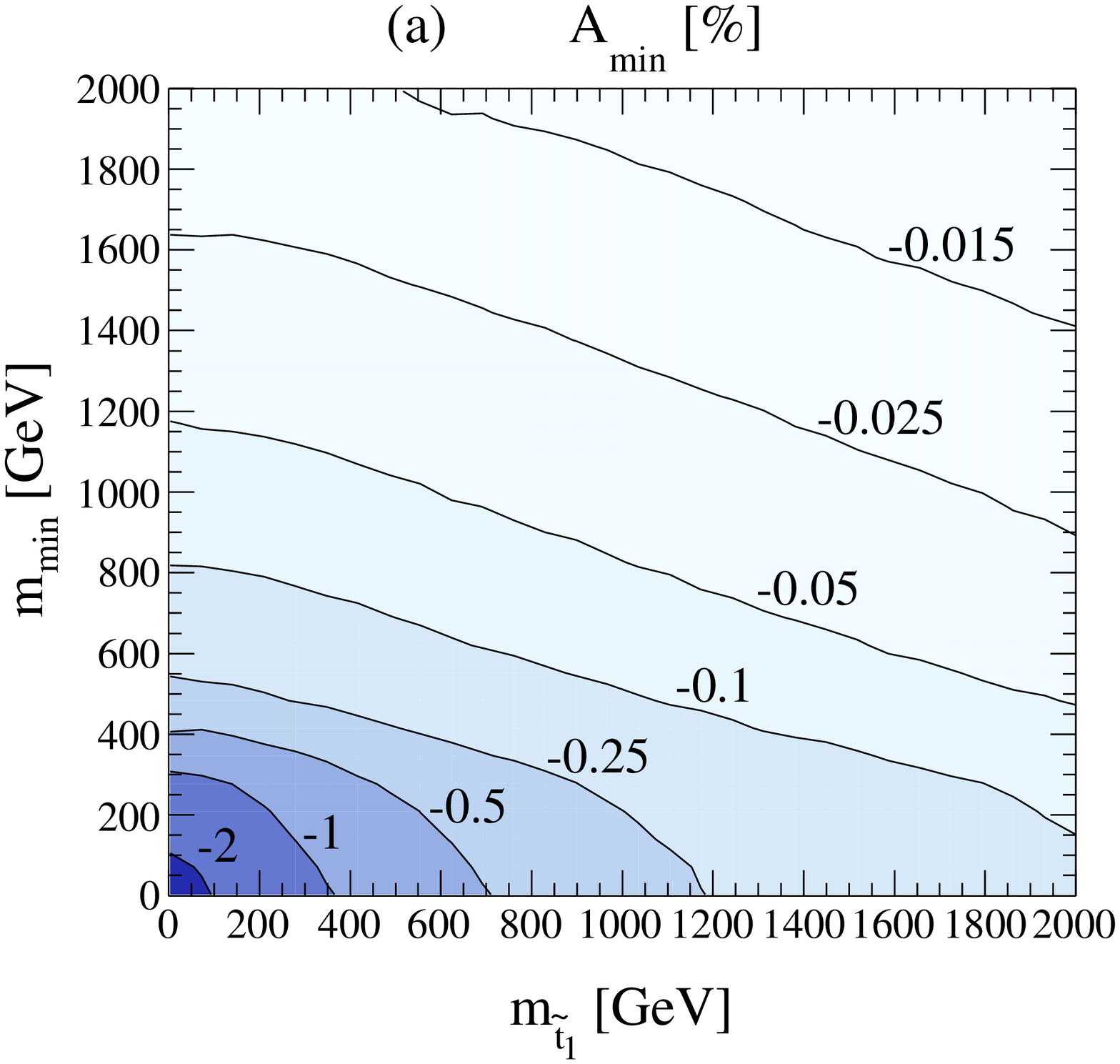}\hspace*{-2mm}
\includegraphics[scale=0.41]{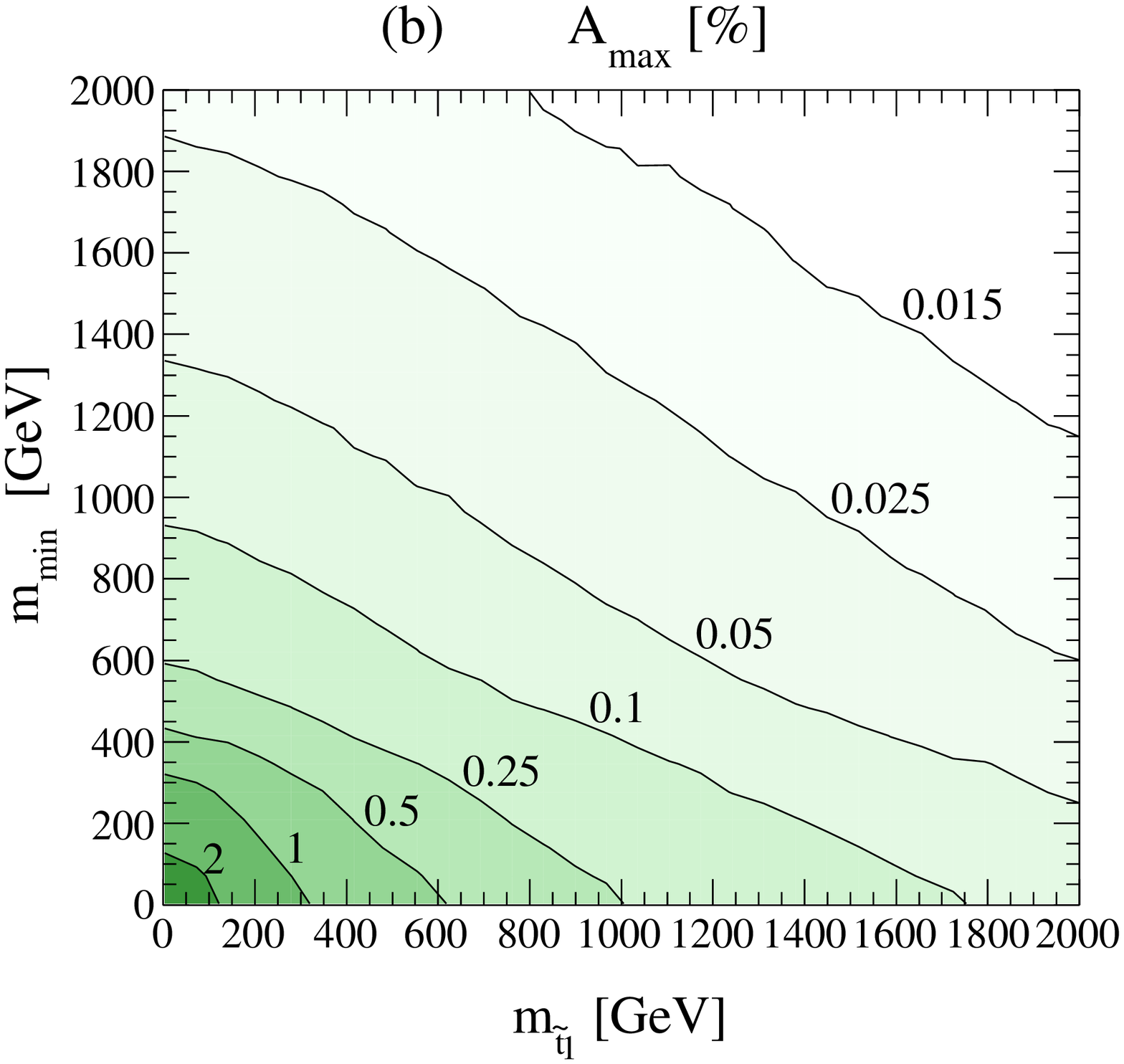}\hspace*{-5mm}\mbox{}\\[-8pt]
\caption{Bounds on the MSSM one-loop contributions to $A_{\text{FB}}^{t\bar
  t}$ with $M_{t\bar{t}}>450$~GeV: (a) Lower and (b) upper bounds on
  $A_{\text{FB}}^{t\bar t}$ (in percent) as functions of the lighter top
  squark mass $m_{\tilde{t}_{1}}$ and $m_{min}$, which is the lightest
  other SUSY particle in the loops except for  $m_{\tilde{t}_2}$, i.~e.
  $m_{min} = \rm{Min}\{
  m_{\tilde{g}},m_{\tilde{\chi}_{1,2,3,4}},m_{\tilde{u}_{1,2}},m_{\tilde{d}_{1,2}}
  \}$.  }
\label{fig:afb_total_mt450} %
\end{figure}

\begin{figure}[htb]
\hspace*{-6mm}\includegraphics[scale=0.41]{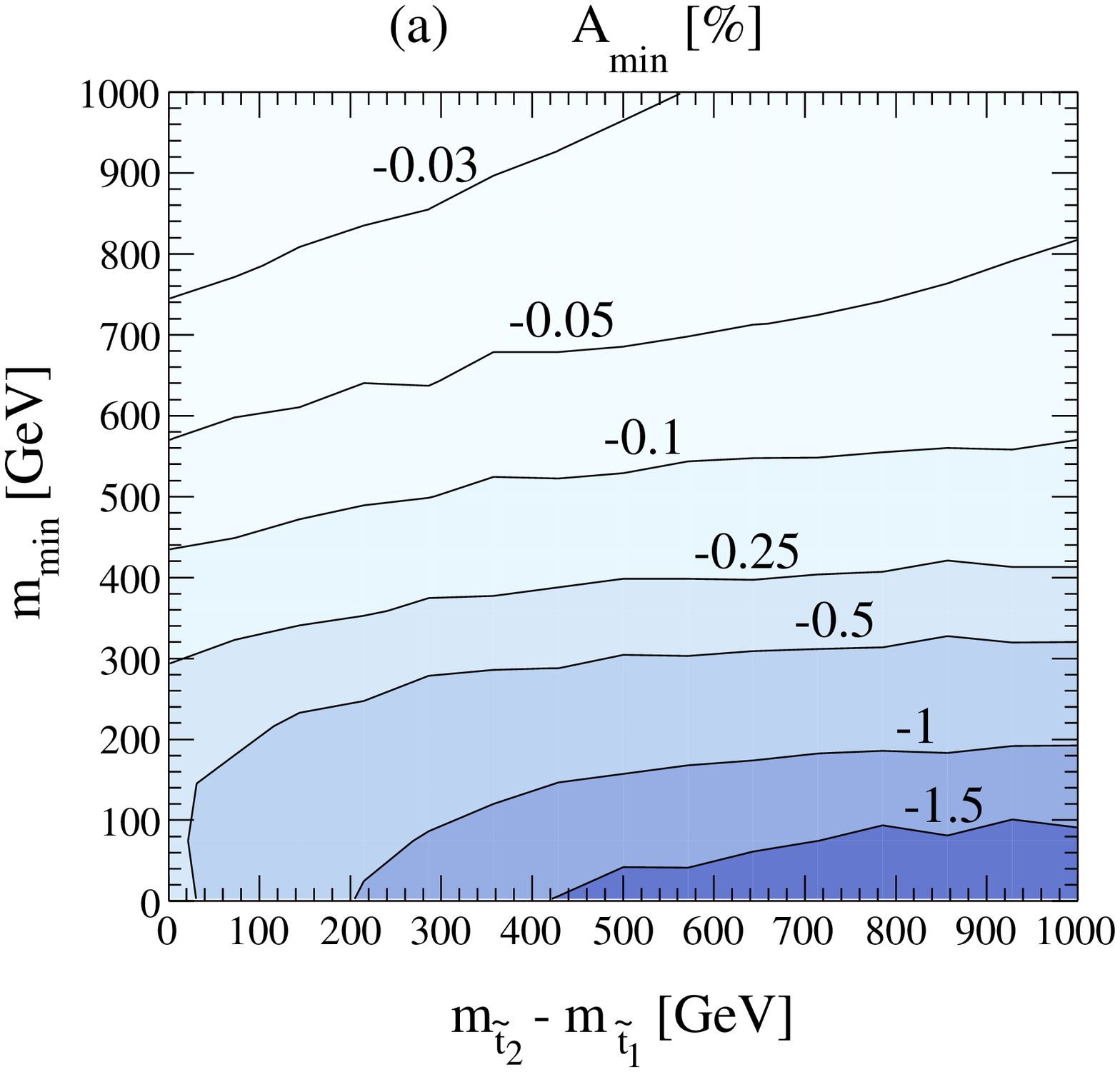}\hspace*{-2mm}\includegraphics[scale=0.41]{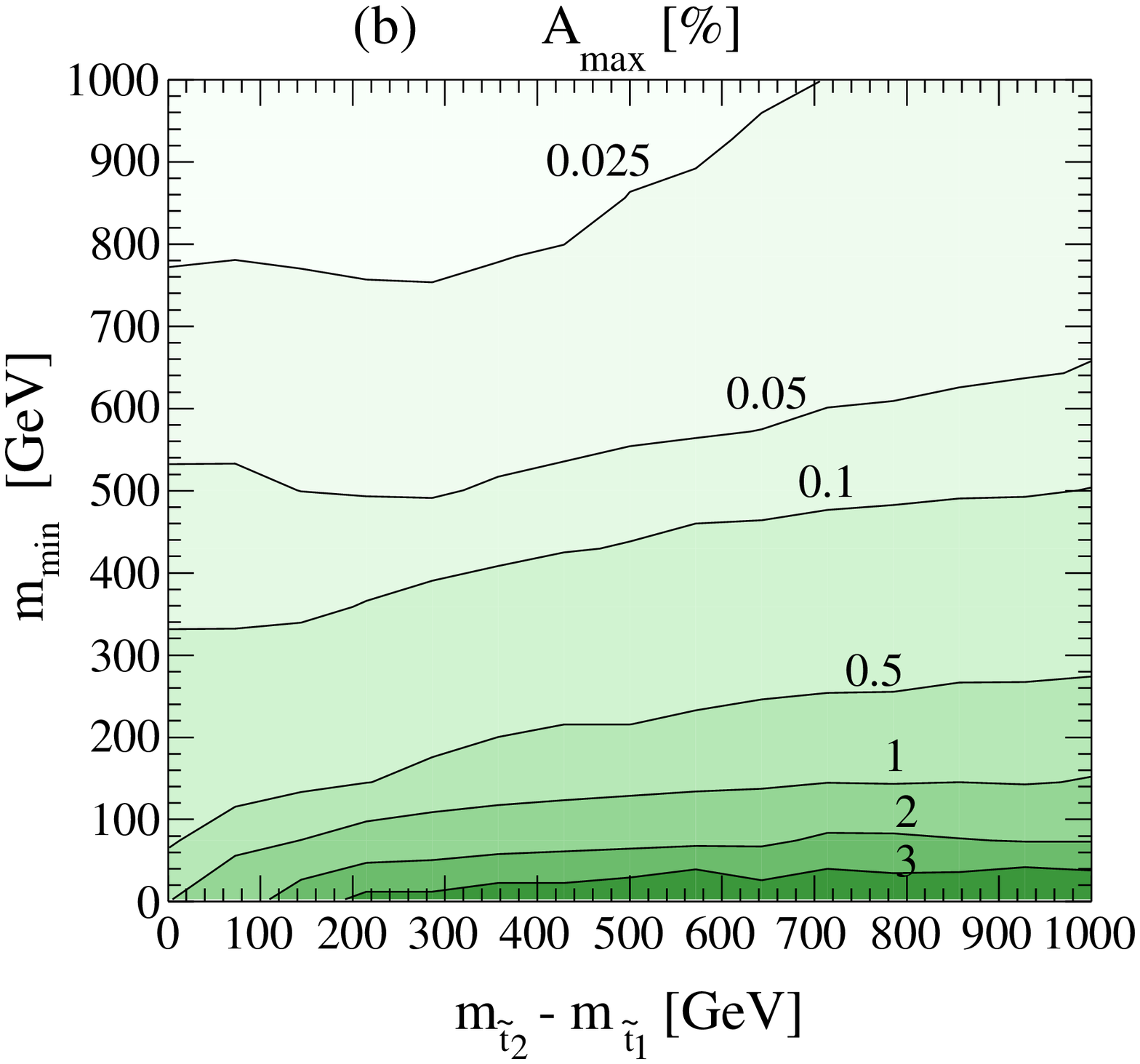}\hspace*{-5mm}\mbox{}\\[-8pt]
\caption{Bounds on the MSSM one-loop contributions to $A_{\text{FB}}^{t\bar
  t}$ with $M_{t\bar{t}}> 450$~GeV: (a) Lower and (b) upper bounds on
  $A_{\text{FB}}^{t\bar t}$ (in percent) as functions of the stop mass
  splitting $m_{\tilde{t}_2} -m_{\tilde{t}_1}$ and $m_{min}$, which is
  the lightest SUSY particle in the  loops, i.~e. $m_{min} = \rm{Min}\{m_{\tilde{t}_{1,2}},
  m_{\tilde{g}},m_{\tilde{\chi}_{1,2,3,4}},m_{\tilde{u}_{1,2}},m_{\tilde{d}_{1,2}}
  \}$.}
\label{fig:afb_mst1mst2_cut} %
\end{figure}

Combining the SUSY QCD and SUSY EW one-loop contributions as well as
taking into account all $q\bar q$-initiated $t\bar t$ production
channels, the lower and upper bounds on $A_{\text{FB}}^{t\bar t}$ in the
complex MFV-MSSM, $A_{min}$ and $A_{max}$, are shown in
Figs.~\ref{fig:afb_total_mt450} and~\ref{fig:afb_mst1mst2_cut}.
In Fig.~\ref{fig:afb_total_mt450} we show the dependence of the bounds
$A_{min,max}$ on the lighter stop quark mass $m_{\tilde{t}_1}$ and on
$m_{min}$, which is the lightest other SUSY particle in the loop
except for  $m_{\tilde{t}_2}$, i.~e. $m_{min} = \rm{Min}\{
m_{\tilde{g}},m_{\tilde{\chi}_{1,2,3,4}},m_{\tilde{u}_{1,2}},m_{\tilde{d}_{1,2}}
\}$. For small masses of $m_{\tilde{t}_1} <200$~GeV
and $m_{min} <200$~GeV, the upper bound on $A_{\text{FB}}^{t\bar t}$ is with
up to $+3\%$ somewhat larger than the absolute value of the lower
bound with $-2\%$. This is because the SUSY QCD bounds,
Eq.~\eqref{eq:sqcdlimits}, are in general not symmetric and in this
region the contribution from $A_{q,ij}^{(3)}$ is non-negligible and
positive as shown in Fig.~\ref{fig:m_afbsqcd}.

In general, one observes that smaller up-squark masses
  $m_{\tilde{u}_1}$ and top squark masses $m_{\tilde{t}_1}$ lead to
  larger MSSM one-loop contributions to the forward-backward
  asymmetry.  The same is valid for the gluino mass when
  $m_{\tilde{g}}>250$~GeV. For $m_{\tilde{g}}<250$~GeV the asymmetry
  is dominated by the contribution of the resonant gluino pair.  Since
  the SUSY EW one-loop corrections are sub-leading, the asymmetry has
  a very small dependence on the neutralino mass.  Furthermore, the
  asymmetry is larger for a larger mass splitting of $m_{\tilde{t}_1}$
  and $m_{\tilde{t}_2}$.  This is illustrated for the lower bound on
  $A_{\text{FB}}^{t\bar t}$ in Fig.~\ref{fig:afb_mst1mst2_cut}(a) and the
  upper bound on $A_{\text{FB}}^{t\bar t}$ in
  Fig.~\ref{fig:afb_mst1mst2_cut}(b), where $A_{min,max}$ are shown in
  dependence of $m_{\tilde{t}_2}-m_{\tilde{t}_1}$ and on $m_{min}$,
  which here is the lightest SUSY particle in the loops,
  i.~e. $m_{min} = \rm{Min}\{m_{\tilde{t}_{1,2}},
  m_{\tilde{g}},m_{\tilde{\chi}_{1,2,3,4}},m_{\tilde{u}_{1,2}},m_{\tilde{d}_{1,2}}
  \}$.

As can be seen, when the stop masses are degenerate, then even for
tiny $m_{min}$ one finds $-0.5\% < A_{\text{FB}}^{t\bar t} < + 1\%$. For a
stop mass splitting of $m_{\tilde{t}_2}-m_{\tilde{t}_1} = 500$~GeV and
$m_{min}>300$~GeV one can obtain bounds up to
$|A_{min,max}|=0.5\%$. Larger values for the asymmetry are possible
when the stop mass splitting is even larger and $m_{min}<300$~GeV.
A similar behavior, but not as pronounced, can be observed for the
up-squark mass splitting $m_{\tilde{u}_2}-m_{\tilde{u}_1}$. For the
bounds shown in Fig.~\ref{fig:afb_mst1mst2_cut}(a) and
Fig.~\ref{fig:afb_mst1mst2_cut}(b),  $\tilde{u}_2$ is basically decoupled
and $m_{\tilde{u}_1}\approx m_{min}$.  In the case when the two
up-squark mass eigenstates are degenerate and $m_{\tilde{u}} \approx
m_{min}>100$~GeV, one can roughly divide the given limits of
Fig.~\ref{fig:afb_mst1mst2_cut} by a factor of two to obtain the limits
for the case of degenerate up-squark masses
$m_{\tilde{u}_1}=m_{\tilde{u}_2}$. For $m_{\tilde{u}} \approx
m_{min}<100$~GeV the limits decrease only mildly.

As discussed earlier,  SUSY searches at the LHC, interpreted in
the CMSSM and Simplified Models, impose stringent limits on the masses
of squarks and gluinos.  An updated summary of the limits obtained by
CMS and ATLAS can be found in
Refs.~\cite{Koay:2012ks,Fehling-Kaschek:1423303}.  In the following we
illustrate the impact of some of these limits on the bounds on
$A_{\text{FB}}^{t\bar t}$.  For instance, ATLAS SUSY searches in events with
jets and missing transverse energy interpreted in the CMSSM find that
gluinos and squarks lighter than about
$950$~GeV~\cite{Aad:2011ib,Fehling-Kaschek:1423303} are ruled out 
for
$m_{\tilde g} = m_{\tilde q}$, while for $m_{\tilde q} \gg m_{\tilde
g}$ the gluino can be somewhat lighter, i.~e. $m_{\tilde
g}>680$~GeV~\cite{Aad:2011qa,Fehling-Kaschek:1423303} for $m_{\tilde
q}=2 m_{\tilde g}$ ( all limits are at $95\%$ C.~L.). 
Note that these squark mass limits do not apply
for the third generation squarks due to slightly different production
mechanisms for stops and sbottoms and the different decay pattern of
top squarks.  As can be seen in Fig.~\ref{fig:afb_total_mt450},
in this scenario with $m_{\tilde g}=m_{min}=680$~GeV the upper bound
on $A_{\text{FB}}^{t\bar t}$ can be at most $A_{max} \approx 0.15\%$ for
$m_{\tilde t_1} \lsim 200$~GeV.

Note that these mass limits depend on the assumptions that go into the
extraction of the limits.  For instance, when $m_{\tilde{g}}\gg
m_{\tilde{q}}$ the SUSY signal is dominated by squark pair production
in the $gg\to\tilde{q}\bar{\tilde{q}}$ channel and thus is
proportional to the number of degenerate squarks.  The signal cross
section drops significantly if all but one squark are decoupled, which
weakens the mass limits obtained under assumptions of degenerate
squarks. The sensitivity of these SUSY searches also drops when the
neutralino mass $m_{\tilde{\chi}_{1}^{0}}$ is increased.  For example,
as shown in Ref.~\cite{Koay:2012ks} the CMS limit on the gluino mass obtained 
in a Simplified Model decreases
from about 850~GeV to 400~GeV, if the neutralino mass is increased
from $0$~GeV to $m_{\tilde g}-200$~GeV.  In this scenario, the upper
bound on $A_{\text{FB}}^{t\bar t}$ increases from about $0.1\%$ ($m_{min}=850$~GeV)  to 
$A_{max} \approx 0.5\%$ ($m_{min}=400$~GeV) for $m_{\tilde t_1} \lsim 200$~GeV, as shown in
Fig.~\ref{fig:afb_total_mt450}.

Moreover, the  extraction of mass limits on squarks and gluinos
gets even more involved if the (light) squarks do not decay directly
into the lightest neutralino.  And finally, one might consider MSSM
scenarios where the neutralino is not the LSP.  The LHC SUSY
searches of
Refs.~\cite{ATLAS-CONF-2011-130,Aad:2011ib,Chatrchyan:2011zy} do not
strictly require a neutralino LSP.  They remain valid as long as
squarks and gluinos decay dominantly into missing energy and one or
two jets, respectively. A different type of analysis is required if
colored SUSY particles are stable or quasi-stable. This can, for
example, happen in gauge mediated SUSY breaking scenarios with a
gravitino LSP and a squark next-to-LSP (NLSP).  Stable or quasi-stable
SUSY particles would form so-called $R$-hadrons with specific detector
signals. A search for $R$-hadrons was presented in
Ref.~\cite{Aad:2011yf} and mass limits of $300$~GeV and $600$~GeV are
given for (quasi-)stable squarks and gluinos, respectively.   Again
with smaller squark/gluino mass, the bounds on $A_{\text{FB}}^{t\bar t}$
increase in this scenario from about $0.25\%$ ($m_{min}=600$~GeV) 
to almost $|A_{min,max}|\approx 1\%$ ($m_{min}=300$~GeV) 
for $m_{\tilde t_1} \lsim 100$~GeV, as shown in
Fig.~\ref{fig:afb_total_mt450}.

As illustrated  on these examples, since our results for the
bounds on the forward-backward asymmetry have been obtained with the
only assumption that we work within the MFV-MSSM, the impact of a
number of specific scenarios on the upper and lower bounds on
$A_{\text{FB}}^{t\bar t}$ can be estimated.

%
\section{Conclusions}\label{sec:conclusions}
%
The study of asymmetries in $t\bar t$ production, such as the
forward-backward charge asymmetry, parity violating asymmetries in
polarized $t\bar t$ production, and spin correlations between the $t$
and $\bar t$, may provide a window to non-SM physics complementary to
direct searches for non-SM particles. A recent measurement of the
corrected (parton-level) forward-backward charge asymmetry in $t\bar
t$ production at the Tevatron deviates from the SM prediction by about
$3\sigma$ in the region $M_{t\bar t}>450$~GeV. Provided the SM
prediction is under good theoretical control, this discrepancy may
leave room for an interpretation as a signal of non-SM physics, and a
number of non-SM scenarios have been proposed in the literature. In
this paper we calculated the SUSY EW and SUSY QCD one-loop corrections
to $A_{\text{FB}}^{t\bar t}$ within the MFV-MSSM and derived general lower
and upper bounds $A_{min},A_{max}$ on $A_{\text{FB}}^{t\bar t}$ at the
Tevatron by performing a comprehensive scan over the relevant MSSM
parameter space. Since the MSSM one-loop corrections to $A_{\text{FB}}^{t\bar
t}$ are dominated by SUSY QCD one-loop corrections, these bounds
strongly depend on the squark and gluino masses and are much less
affected by the neutralino mass.  As expected, these loop effects are
more pronounced for smaller sparticle masses in the loops and larger
stop/squark mass splittings.  For very small masses, $m_{\tilde{t}_1}
<200$~GeV and $m_{min} = \rm{Min}\{
m_{\tilde{g}},m_{\tilde{u}_{1,2}},m_{\tilde{d}_{1,2}}\}<200$~GeV, we
find $-2\% < A_{\text{FB}}^{t\bar t}<+3\%$, which is comparable in size to
the SM EW corrections to $A_{\text{FB}}^{t\bar t}$.  This is the maximum
possible SUSY loop-induced asymmetry that can be obtained within the
MFV-MSSM at the Tevatron, with $M_{t\bar t}>450$~GeV.  It is important
to emphasize that apart from working within the MFV-MSSM, no
additional assumptions or constraints have been imposed in our
derivation of the bounds on $A_{\text{FB}}^{t\bar t}$ and that from the
general bounds presented in this paper an estimate can be obtained of
how they change when assuming a specific SUSY scenario.  For example,
to illustrate the impact of squark and gluino mass limits obtained
within the CMSSM at the LHC, we obtain from the results presented in
Fig.~\ref{fig:afb_total_mt450}  
that for gluino and squark masses
in the range 850~GeV-1000~GeV, $A_{max} \approx 0.1\%$ for a light top squark of
$m_{\tilde t_1} = 200$~GeV and $m_{min}=850$~GeV, and $A_{max}
\approx 0.05\%$ for $m_{\tilde t_1} = 600$~GeV and
$m_{min}=1$~TeV.

\section*{Acknowledgments}

The work of S.~B. is supported by the Initiative and Networking Fund
of the Helmholtz Association, contract HA-101 (`Physics at the
Terascale') and by the Research Center `Elementary Forces and
Mathematical Foundations' of the Johannes-Gutenberg-Universit\"at
Mainz.  The work of D.~W. is supported by the National Science
Foundation under grant No.~NSF-PHY-0547564 and No.~NSF-PHY-0757691.
The work of M.~W. is partially supported by project DFG NI 1105/2-1.

%
\section{Appendix}
%

\subsection{Gluino-squark-quark couplings}\label{sec:gsqcoup}

The gluino-squark-quark couplings as defined in
Eq.~\eqref{eq:Def_Gamma_n} are~\cite{Hahn:2001rv}
\begin{eqnarray}
  g_{n}^{\pm} & = & i\sqrt{2}g_{s}\,\hat{g}_{n}^{\,\pm}\qquad{\rm with}\label{eq:def_g_gluino}\\
  \hat{g}_{1}^{\,\lambda} & = & \lambda U_{i,\lambda}^{\tilde{q}}e^{i\lambda\phi/2}\quad,\qquad\hat{g}_{4}^{\,\lambda}=\lambda U_{j,\lambda}^{\tilde{t}}e^{i\lambda\phi/2}\quad,\qquad(\lambda=+,-)\,\,.\label{eq:def_hat_g_gluino}\end{eqnarray}
The index $i$ and $j$ have been omitted in the definition of $g_{n}^{\pm}$
of Eq.~\eqref{eq:def_g_gluino} to avoid large chains of indices in
Section~\ref{sec:MSSM-Contributions-to_Afb}. The index $i$ always
refers to the squark index of flavor $q=\{u,d,s,c,b\}$ of the vertices
$\Gamma_{1}$ and $\Gamma_{2}$ and the index $j$ always refers to
the stop quark index of $\Gamma_{3}$ and $\Gamma_{4}$. The complex
phase of the gluino mass $M_{3}$ is denoted by $\phi$.
The couplings are related by
\[
\hat{g}_{2}^{\,\pm}=\hat{g}_{1}^{\,\mp*}\quad,\qquad\hat{g}_{3}^{\,\pm}=\hat{g}_{4}^{\,\mp*}.\]
The unitary squark mixing matrices are given as
\begin{equation}
U^{\tilde{q}}=\begin{pmatrix}U_{1,-}^{\tilde{q}} & U_{1,+}^{\tilde{q}}\\
U_{2,-}^{\tilde{q}} & U_{2,+}^{\tilde{q}}\end{pmatrix}\quad,\qquad U^{\tilde{t}}=\begin{pmatrix}U_{1,-}^{\tilde{t}} & U_{1,+}^{\tilde{t}}\\
U_{2,-}^{\tilde{t}} &
U_{2,+}^{\tilde{t}}\end{pmatrix}\label{eq:squarkmix}\end{equation}
For the squark mass and mixing matrices we use the conventions of Ref.~\cite{Hahn:2001rv}.

\subsection{Neutralino-squark-quark couplings }\label{sec:nsqcoup}
 
For the neutralino-squark-quark couplings, we again follow the
notation of Ref.~\cite{Hahn:2001rv} , where explicit expressions for
these couplings can be found.  For completeness, since the $u\bar
u$-channel is the dominant $t\bar t$ production process, we provide
here the neutralino-up-quark-squark coupling, which reads with the
restriction $m_{u}=0$~GeV~\cite{Hahn:2001rv}:
\noindent \begin{eqnarray}
g_{n}^{\,\pm} & = & i\sqrt{2}\, e\,\hat{g}_{n}^{\,\pm}\qquad{\rm with}\label{eq:def_g_Neut}\\
\hat{g}_{1}^{\,+} & = & \frac{2}{3c_{W}}\cdot N_{k,1}\cdot U_{i,+}^{\tilde{u}}\nonumber \\
\hat{g}_{1}^{\,-} & = & {}-\frac{1}{2s_{W}}\cdot N_{k,2}^{*}\cdot U_{i,-}^{\tilde{u}}-\frac{1}{6c_{W}}\cdot N_{k,1}^{*}\cdot U_{i,-}^{\tilde{u}}\nonumber \\
\hat{g}_{4}^{\,+} & = & \frac{1}{6c_{W}s_{W}m_{W}s_{\beta}}\cdot\left[4m_{W}s_{\beta}s_{W}N_{k,1}\cdot U_{j,+}^{\tilde{t}}-3c_{W}m_{t}N_{k,4}\cdot U_{j,-}^{\tilde{t}}\right]\nonumber \\
\hat{g}_{4}^{\,-} & = &
\frac{{}-1}{6c_{W}s_{W}m_{W}s_{\beta}}\cdot\left[\left(3c_{W}m_{W}s_{\beta}N_{k,2}^{*}+s_{W}m_{W}s_{\beta}N_{k,1}^{*}\right)\cdot
  U_{j,-}^{\tilde{t}}+3c_{W}m_{t}N_{k,4}^{*}\cdot U_{j,+}^{\tilde{t}}\right]
\label{eq:def_hat_g_Neut} 
\end{eqnarray}
where we used the shorthand notations
$c_W=\cos\theta_W,s_W=\sin\theta_W$ and
$s_\beta=\sin\beta$ with $\tan\beta=\frac{v_u}{v_d}$ the ratio of the two
Higgs field vacuum expectation values. 
Here the coupling parameters are related by
\[g_{2}^{\pm}=g_{1}^{\mp*}\quad,\qquad g_{3}^{\pm}=g_{4}^{\mp*}.\]
The neutralino mass matrices in the used convention have been taken
from Ref.~\cite{Frank:2006yh}.
%
\subsection{Analytic Expressions for the one-loop functions $D^a$}\label{sec:da}
%
The partonic differential cross section
$\frac{d\hat{\sigma}^{(a,b)}}{d\cos\theta}$ for the direct box diagrams of
Fig.~\ref{fig:boxes}(a) and crossed box of Fig.~\ref{fig:boxes}(b) are given
in Eq.~\eqref{eq:intf_a} and Eq.~\eqref{eq:intf_b}, respectively, in terms of coupling parameters
and loop functions $D^{a}$.  Neglecting the initial-state quark masses and
using the mass assignments of Fig.~\ref{fig:boxes}, the functions
$D^{a}(\hat{s},\cos\theta)$ are given with
$\hat{t}=m_{t}^{2}-\frac{\hat{s}}{2}\left(1-\beta_t\cos\theta\right)$ as
\begin{eqnarray}
D^{+-++}(\hat{s},\cos\theta) & = & -\frac{1}{32\pi^2\hat{s}}\cdot\left[-m_{2}m_{t}\hat{s}^{2}D_{1}+\left(-m_{t}^{4}-m_{t}^{2}\hat{s}+2m_{t}^{2}\hat{t}-\hat{t}^{2}\right)m_{t}m_{2}D_{2}\right]\nonumber \\
D^{-+++}(\hat{s},\cos\theta) & = & -\frac{1}{32\pi^2\hat{s}}\cdot m_{t}m_{4}\left[\left(-m_{t}^{4}-m_{t}^{2}\hat{s}+2m_{t}^{2}\hat{t}-\hat{t}^{2}\right)D_{2}-\hat{s}^{2}D_{3}\right]\nonumber \\
D^{-++-}(\hat{s},\cos\theta) & = & -\frac{1}{32\pi^2\hat{s}}\cdot\left\{ D_{00}\cdot\left(-2m_{t}^{4}+2m_{t}^{2}\hat{s}+4m_{t}^{2}\hat{t}-2\hat{s}^{2}-4\hat{s}\hat{t}-2\hat{t}^{2}\right)\right.\nonumber \\
 &  & \qquad\quad{}+m_{t}^{2}\hat{s}^{2}D_{12}+\left(m_{t}^{4}\hat{s}-m_{t}^{2}\hat{s}^{2}-2m_{t}^{2}\hat{s}\hat{t}+\hat{s}^{3}+2\hat{s}^{2}\hat{t}+\hat{s}\hat{t}^{2}\right)D_{13}\nonumber \\
 &  & \qquad\quad\left.{}+\left(m_{t}^{6}+m_{t}^{4}\hat{s}-2m_{t}^{4}\hat{t}+m_{t}^{2}\cdot\hat{t}^{2}\right)D_{22}+m_{t}^{2}\hat{s}^{2}D_{23}\right\} \nonumber \\
D^{-+-+}(\hat{s},\cos\theta) & = & -\frac{1}{32\pi^2\hat{s}}\cdot
m_{2}m_{4}\cdot\left(m_{t}^{4}+m_{t}^{2}\hat{s}-2m_{t}^{2}\hat{t}+\hat{t}^{2}\right)D_{0},\label{eq:Definition_of_Da}
\end{eqnarray}
 where
$D_{i,ij}=D_{i,ij}\left(0,m_{t}^{2},m_{t}^{2},0,\hat{t},\hat{s},m_{1}^{2},m_{2}^{2},m_{3}^{2},m_{4}^{2}\right)$
are written in the convention of Ref.~\cite{Hahn:1998yk}.  From these
expressions the contribution to Fig.~\ref{fig:boxes}(b) can be
obtained by replacing $D^{a}(\hat{s},\cos\theta)\to
D^{a}(\hat{s},-\cos\theta)$ and multiplying by a factor of $\,(-1)$
for exchanging the final-state fermions.

\bibliographystyle{unsrtnat}
\bibliography{bibpptt}

\begin{thebibliography}{63}
\expandafter\ifx\csname natexlab\endcsname\relax\def\natexlab#1{#1}\fi

\bibitem[Aaltonen et~al.(2008)]{Aaltonen:2008hc}
Aaltonen, T. and others, {\em Phys.Rev.Lett.}
\newblock 101 \penalty0 (2008) \penalty0 202001 [0806.2472].

\bibitem[Aaltonen et~al.(2011)]{Aaltonen:2011kc}
Aaltonen, T. and others, {\em Phys.Rev.}
\newblock D83 \penalty0 (2011) \penalty0 112003 [1101.0034].

\bibitem[Collaboration(2011)]{unknown:2011comb}
The CDF Collaboration.
\newblock  \penalty0 (2011) [CDF Conf. Note 10584].

\bibitem[Abazov et~al.(2008)]{:2007qb}
Abazov, V.M. and others, {\em Phys.Rev.Lett.}
\newblock 100 \penalty0 (2008) \penalty0 142002 [0712.0851].

\bibitem[Abazov et~al.(2011)]{Abazov:2011rq}
Abazov, Victor Mukhamedovich and others, {\em Phys. Rev.}
\newblock D84 \penalty0 (2011) \penalty0 112005 [1107.4995].

\bibitem[K{\"u}hn and Rodrigo(1999)]{Kuhn:1998kw}
K{\"u}hn, Johann H. and Rodrigo, German, {\em Phys. Rev.}
\newblock D59 \penalty0 (1999) \penalty0 054017 [hep-ph/9807420].

\bibitem[K{\"u}hn and Rodrigo(1998)]{Kuhn:1998jr}
K{\"u}hn, Johann H. and Rodrigo, German, {\em Phys. Rev. Lett.}
\newblock 81 \penalty0 (1998) \penalty0 49--52 [hep-ph/9802268].

\bibitem[Bowen et~al.(2006)Bowen, Ellis, and Rainwater]{Bowen:2005ap}
Bowen, M. T. and Ellis, S. D. and Rainwater, D., {\em Phys. Rev.}
\newblock D73 \penalty0 (2006) \penalty0 014008 [hep-ph/0509267].

\bibitem[Antunano et~al.(2008)Antunano, K{\"u}hn, and Rodrigo]{Antunano:2007da}
Antunano, Oscar and K{\"u}hn, Johann H. and Rodrigo, German, {\em Phys. Rev.}
\newblock D77 \penalty0 (2008) \penalty0 014003 [0709.1652].

\bibitem[Almeida et~al.(2008)Almeida, Sterman, and Vogelsang]{Almeida:2008ug}
Almeida, Leandro G. and Sterman, George F. and Vogelsang, Werner, {\em Phys.
  Rev.}
\newblock D78 \penalty0 (2008) \penalty0 014008 [0805.1885].

\bibitem[Ahrens et~al.(2011)Ahrens, Ferroglia, Neubert, Pecjak, and
  Yang]{Ahrens:2011uf}
Ahrens, Valentin and Ferroglia, Andrea and Neubert, Matthias and Pecjak, Ben D.
  and Yang, Li Lin, {\em Phys. Rev.}
\newblock D84 \penalty0 (2011) \penalty0 074004 [1106.6051].

\bibitem[Hollik and Pagani(2011)]{Hollik:2011ps}
Hollik, Wolfgang and Pagani, Davide, {\em Phys. Rev.}
\newblock D84 \penalty0 (2011) \penalty0 093003 [1107.2606].

\bibitem[K{\"u}hn and Rodrigo(2012)]{Kuhn:2011ri}
K{\"u}hn, Johann H. and Rodrigo, German, {\em JHEP}.
\newblock 01 \penalty0 (2012) \penalty0 063 [1109.6830].

\bibitem[Manohar and Trott(2012)]{Manohar:2012rs}
Manohar, Aneesh V. and Trott, Michael.
\newblock  \penalty0 (2012) [1201.3926].

\bibitem[Kidonakis(2011)]{Kidonakis:2011zn}
Kidonakis, Nikolaos, {\em Phys. Rev.}
\newblock D84 \penalty0 (2011) \penalty0 011504 [1105.5167].

\bibitem[Aguilar-Saavedra and Perez-Victoria(2011)]{AguilarSaavedra:2011ug}
Aguilar-Saavedra, J. A. and Perez-Victoria, M., {\em JHEP.}
\newblock 1109 \penalty0 (2011) \penalty0 097 [1107.0841].

\bibitem[Davoudiasl et~al.(2011)Davoudiasl, McElmurry, and
  Soni]{Davoudiasl:2011tv}
Davoudiasl, Hooman and McElmurry, Thomas and Soni, Amarjit.
\newblock  \penalty0 (2011) [1108.1173].

\bibitem[Cui et~al.(2011)Cui, Han, and Schwartz]{Cui:2011xy}
Cui, Yanou and Han, Zhenyu and Schwartz, Matthew D., {\em JHEP}.
\newblock 07 \penalty0 (2011) \penalty0 127 [1106.3086].

\bibitem[Isidori and Kamenik(2011)]{Isidori:2011dp}
Isidori, Gino and Kamenik, Jernej F., {\em Phys. Lett.}
\newblock B700 \penalty0 (2011) \penalty0 145--149 [1103.0016].

\bibitem[Nilles(1984)]{Nilles:1983ge}
Nilles, Hans Peter, {\em Phys. Rept.}
\newblock 110 \penalty0 (1984) \penalty0 1.

\bibitem[Haber and Kane(1985)]{Haber:1984rc}
Haber, Howard E. and Kane, Gordon L., {\em Phys. Rept.}
\newblock 117 \penalty0 (1985) \penalty0 75.

\bibitem[Buras et~al.(2001)Buras, Gambino, Gorbahn, Jager, and
  Silvestrini]{Buras:2000dm}
Buras, A.J. and Gambino, P. and Gorbahn, M. and Jager, S. and Silvestrini, L.,
  {\em Phys.Lett.}
\newblock B500 \penalty0 (2001) \penalty0 161--167 [hep-ph/0007085].

\bibitem[D'Ambrosio et~al.(2002)D'Ambrosio, Giudice, Isidori, and
  Strumia]{D'Ambrosio:2002ex}
D'Ambrosio, G. and Giudice, G.F. and Isidori, G. and Strumia, A., {\em
  Nucl.Phys.}
\newblock B645 \penalty0 (2002) \penalty0 155--187 [hep-ph/0207036].

\bibitem[Li et~al.(1995)Li, Hu, Yang, and Hu]{Li:1995fj}
Li, Chong-Sheng and Hu, Bing-Quan and Yang, Jin-Min and Hu, Chen-Guo, {\em
  Phys. Rev.}
\newblock D52 \penalty0 (1995) \penalty0 5014--5017.

\bibitem[Alam et~al.(1997)Alam, Hagiwara, and Matsumoto]{Alam:1996mh}
Alam, S. and Hagiwara, K. and Matsumoto, S., {\em Phys. Rev.}
\newblock D55 \penalty0 (1997) \penalty0 1307--1315 [hep-ph/9607466].

\bibitem[Sullivan(1997)]{Sullivan:1996ry}
Sullivan, Zack, {\em Phys. Rev.}
\newblock D56 \penalty0 (1997) \penalty0 451--457 [hep-ph/9611302].

\bibitem[Zhou and Li(1997)]{Zhou:1997fw}
Zhou, Hong-Yi and Li, Chong-Sheng, {\em Phys. Rev.}
\newblock D55 \penalty0 (1997) \penalty0 4421--4429.

\bibitem[Yu et~al.(1999)Yu, Pietschmann, Ma, Han, and Yi]{Yu:1998xv}
Yu, Zeng-Hui and Pietschmann, H. and Ma, Wen-Gan and Han, Liang and Yi, Jiang,
  {\em Eur. Phys. J.}
\newblock C9 \penalty0 (1999) \penalty0 463--477 [hep-ph/9804331].

\bibitem[Wackeroth(1998)]{Wackeroth:1998wm}
Wackeroth, D.
\newblock  \penalty0 (1998) [hep-ph/9807558].

\bibitem[Berge et~al.(2007)Berge, Hollik, Mosle, and Wackeroth]{Berge:2007dz}
Berge, Stefan and Hollik, Wolfgang and Mosle, Wolf M. and Wackeroth, Doreen,
  {\em Phys. Rev.}
\newblock D76 \penalty0 (2007) \penalty0 034016 [hep-ph/0703016].

\bibitem[Ross and Wiebusch(2007)]{Ross:2007ez}
Ross, D. A. and Wiebusch, M., {\em JHEP}.
\newblock 11 \penalty0 (2007) \penalty0 041 [0707.4402].

\bibitem[Yang and Li(1995)]{Yang:1995hq}
Yang, Jin-Min and Li, Chong-Sheng, {\em Phys. Rev.}
\newblock D52 \penalty0 (1995) \penalty0 1541--1545.

\bibitem[Yang and Li(1996)]{Yang:1996dm}
Yang, Jin Min and Li, Chong Sheng, {\em Phys. Rev.}
\newblock D54 \penalty0 (1996) \penalty0 4380--4384 [hep-ph/9603442].

\bibitem[Kim et~al.(1996)Kim, Lopez, Nanopoulos, and Rangarajan]{Kim:1996nz}
Kim, Jaewan and Lopez, Jorge L. and Nanopoulos, D. V. and Rangarajan, R., {\em
  Phys. Rev.}
\newblock D54 \penalty0 (1996) \penalty0 4364--4373 [hep-ph/9605419].

\bibitem[Hollik et~al.(1998)Hollik, Mosle, and Wackeroth]{Hollik:1997hm}
Hollik, W. and Mosle, W. M. and Wackeroth, D., {\em Nucl. Phys.}
\newblock B516 \penalty0 (1998) \penalty0 29--54 [hep-ph/9706218].

\bibitem[Denner et~al.(1992)Denner, Eck, Hahn, and K{\"u}blbeck]{Denner:1992vz}
Denner, A. and Eck, H. and Hahn, O. and K{\"u}blbeck, J., {\em Nucl. Phys.}
\newblock B387 \penalty0 (1992) \penalty0 467--484.

\bibitem[K{\"u}blbeck et~al.(1990)K{\"u}blbeck, B{\"o}hm, and
  Denner]{Kublbeck:1990xc}
K{\"u}blbeck, J. and B{\"o}hm, M. and Denner, A., {\em Comput. Phys. Commun.}
\newblock 60 \penalty0 (1990) \penalty0 165--180.

\bibitem[Hahn(2001)]{Hahn:2000kx}
Hahn, Thomas, {\em Comput. Phys. Commun.}
\newblock 140 \penalty0 (2001) \penalty0 418--431 [hep-ph/0012260].

\bibitem[Hahn and Schappacher(2002)]{Hahn:2001rv}
Hahn, Thomas and Schappacher, Christian, {\em Comput. Phys. Commun.}
\newblock 143 \penalty0 (2002) \penalty0 54--68 [hep-ph/0105349].

\bibitem[Passarino and Veltman(1979)]{Passarino:1979jh}
Passarino, G. and Veltman, M. J. G., {\em Nucl. Phys.}
\newblock B160 \penalty0 (1979) \penalty0 151.

\bibitem[Vermaseren(2000)]{Vermaseren:2000nd}
Vermaseren, J.A.M.
\newblock  \penalty0 (2000) [math-ph/0010025].

\bibitem[Pumplin et~al.(2002)]{Pumplin:2002vw}
Pumplin, J. and others, {\em JHEP}.
\newblock 07 \penalty0 (2002) \penalty0 012 [hep-ph/0201195].

\bibitem[Hahn and Perez-Victoria(1999)]{Hahn:1998yk}
Hahn, T. and Perez-Victoria, M., {\em Comput.Phys.Commun.}
\newblock 118 \penalty0 (1999) \penalty0 153--165 [hep-ph/9807565].

\bibitem[Nakamura et~al.(2010)]{Nakamura:2010zzi}
Nakamura, K. and others, {\em J. Phys.}
\newblock G37 \penalty0 (2010) \penalty0 075021.

\bibitem[Chatrchyan et~al.(2011)]{Chatrchyan:2011zy}
Chatrchyan, Serguei and others, {\em Phys. Rev. Lett.}
\newblock 107 \penalty0 (2011) \penalty0 221804 [1109.2352].

\bibitem[Chatrchyan et~al.(2012)]{Chatrchyan:2011ek}
Chatrchyan, Serguei and others, {\em Phys. Rev.}
\newblock D85 \penalty0 (2012) \penalty0 012004 [1107.1279].

\bibitem[Khachatryan et~al.(2011)]{Khachatryan:2011tk}
Khachatryan, Vardan and others, {\em Phys. Lett.}
\newblock B698 \penalty0 (2011) \penalty0 196--218 [1101.1628].

\bibitem[Aad et~al.(2011{\natexlab{a}})]{Aad:2011ib}
Aad, Georges and others.
\newblock  \penalty0 (2011) [1109.6572].

\bibitem[Aad et~al.(2012)]{ATLAS:2011ad}
Aad, Georges and others, {\em Phys. Rev.}
\newblock D85 \penalty0 (2012) \penalty0 012006 [1109.6606].

\bibitem[Aad et~al.(2011{\natexlab{b}})]{Aad:2011hh}
Aad, Georges and others, {\em Phys. Rev. Lett.}
\newblock 106 \penalty0 (2011) \penalty0 131802 [1102.2357].

\bibitem[da~Costa et~al.(2011)]{daCosta:2011qk}
da Costa, Joao Barreiro Guimaraes and others, {\em Phys. Lett.}
\newblock B701 \penalty0 (2011) \penalty0 186--203 [1102.5290].

\bibitem[Landau(1959)]{Landau:1959fi}
Landau, L. D., {\em Nucl. Phys.}
\newblock 13 \penalty0 (1959) \penalty0 181--192.

\bibitem[Djouadi et~al.(2007)Djouadi, Kneur, and Moultaka]{Djouadi:2002ze}
Djouadi, Abdelhak and Kneur, Jean-Loic and Moultaka, Gilbert, {\em
  Comput.Phys.Commun.}
\newblock 176 \penalty0 (2007) \penalty0 426--455 [hep-ph/0211331].

\bibitem[DVE(????)]{DVEGAS}
{\em \url{http://hepsource.sf.net/dvegas}}.

\bibitem[Kauer and Zeppenfeld(2002)]{Kauer:2001sp}
Kauer, N. and Zeppenfeld, D., {\em Phys.Rev.}
\newblock D65 \penalty0 (2002) \penalty0 014021 [hep-ph/0107181].

\bibitem[Kauer(2003)]{Kauer:2002sn}
Kauer, N., {\em Phys.Rev.}
\newblock D67 \penalty0 (2003) \penalty0 054013 [hep-ph/0212091].

\bibitem[Brein(2005)]{Brein:2004kh}
Brein, Oliver, {\em Comput. Phys. Commun.}
\newblock 170 \penalty0 (2005) \penalty0 42--48 [hep-ph/0407340].

\bibitem[Koay and Collaboration(2012)]{Koay:2012ks}
Koay, S.A. and Collaboration, CMS.
\newblock  \penalty0 (2012) [1202.1000].

\bibitem[Fehling-Kaschek(2012)]{Fehling-Kaschek:1423303}
M~Fehling-Kaschek.
\newblock Supersymmetry searches at atlas.
\newblock Technical Report ATL-PHYS-PROC-2012-039, CERN, Geneva, Feb 2012.

\bibitem[Aad et~al.(2011{\natexlab{c}})]{Aad:2011qa}
Aad, Georges and others, {\em JHEP}.
\newblock 11 \penalty0 (2011) \penalty0 099 [1110.2299].

\bibitem[ATL(2011)]{ATLAS-CONF-2011-130}
Search for supersymmetry in pp 1 collisions at sqrt(s) = 7 tev in final states
  with missing transverse momentum, b-jets and one lepton with the atlas
  detector.
\newblock Technical Report ATLAS-CONF-2011-130, CERN, Geneva, Sep 2011.

\bibitem[Aad et~al.(2011{\natexlab{d}})]{Aad:2011yf}
Aad, Georges and others, {\em Phys.Lett.}
\newblock B701 \penalty0 (2011) \penalty0 1--19 [1103.1984].

\bibitem[Frank et~al.(2007)Frank, Hahn, Heinemeyer, Hollik, Rzehak,
  et~al.]{Frank:2006yh}
Frank, M. and Hahn, T. and Heinemeyer, S. and Hollik, W. and Rzehak, H. and
  others, {\em JHEP}.
\newblock 0702 \penalty0 (2007) \penalty0 047 [hep-ph/0611326].

\end{thebibliography}

\end{document}